\documentclass[a4paper,11pt]{article}
\pdfoutput=1
\usepackage{jheppub}
\usepackage[utf8]{inputenc}

\usepackage{amsmath}
\usepackage{amssymb}
\usepackage{amsfonts}
\usepackage{graphicx}
\usepackage{xcolor}
\usepackage[labelformat=simple]{subcaption}

\usepackage[english]{babel}
\usepackage{cleveref}
\crefformat{section}{Section~#2#1#3}
\crefformat{figure}{Fig.~#2#1#3}
\crefformat{table}{Table~#2#1#3}
\crefformat{equation}{Eq.~(#2#1#3)}
\crefrangeformat{equation}{Eqs.~(#3#1#4) to~(#5#2#6)}
\crefmultiformat{equation}{Eqs.~(#2#1#3)}{ and~(#2#1#3)}{, (#2#1#3)}{ and~(#2#1#3)}
\crefrangemultiformat{equation}{Eqs.~(#2#1#3)}{ and~(#2#1#3)}{, (#2#1#3)}{ and~(#2#1#3)}
\crefname{appsec}{Appendix}{Appendices}
\newcommand{\be}{\begin{equation}}
\newcommand{\ee}{\end{equation}}

\newcommand{\Gam}{\Gamma}
\newcommand{\eps}{\epsilon}
\newcommand{\pbar}{\bar{p}}

\title{Triple-real contribution to the quark beam function in QCD at next-to-next-to-next-to-leading order}

\author[a]{K. Melnikov,}
\author[a]{R. Rietkerk,}
\author[b]{L. Tancredi}
\author[c]{and C. Wever}

\affiliation[a]{Institute for Theoretical Particle Physics, KIT, Karlsruhe, Germany}
\affiliation[b]{Theoretical Physics Department, CERN, 1211 Geneva 23, Switzerland}
\affiliation[c]{Physik-Department T31, Technische Universität München, James-Franck-Straße 1, D-85748 Garching, Germany}

\emailAdd{kirill.melnikov@kit.edu}
\emailAdd{robbert.rietkerk@kit.edu}
\emailAdd{lorenzo.tancredi@cern.ch}
\emailAdd{christopher.wever@tum.de}

\abstract{We compute the three-loop master integrals required for the calculation of the triple-real contribution to the N$^3$LO quark beam function due to the splitting of a quark into a virtual quark and three collinear gluons, $q \to q^*+ggg$. This provides an important ingredient for the calculation of the leading-color contribution to the quark beam function at N$^3$LO.}


\arxivnumber{1904.02433}

\begin{document} 
\maketitle
\flushbottom

\allowdisplaybreaks

\section{Introduction } 
\label{sec:intro}

The detailed exploration of perturbative Quantum Field Theory has played 
an important role in collider physics during  the last decade.
In fact, the need to study the recently discovered Higgs boson~\cite{Aad:2012tfa,Chatrchyan:2012xdj} and the absence
of any sign of physics beyond the Standard Model in LHC experiments 
are behind  an impressive effort of particle theorists to provide predictions for important LHC observables with high  precision.

Although precision physics at hadron colliders is very difficult, the LHC experiments have been performing very well, having already delivered measurements for multiple observables at the percent level and even beyond, see e.g. Refs.~\cite{Aad:2016naf,Chatrchyan:2014mua,Aaboud:2018hip,Sirunyan:2018goh,Aaboud:2019gxl,Sirunyan:2019bez,Aaboud:2019pkc}. 
Comparing these experimental results with equally precise theoretical predictions, will make it possible to search for New Physics indirectly by probing energy scales far beyond the direct reach of the LHC.

These considerations, augmented by an impressive experimental progress, have been continuously pushing the default standard for theoretical predictions for LHC physics from leading to next-to-leading \cite{Kubar:1980zv,Nason:1987xz,Ohnemus:1991gb,Frixione:1992pj,Ossola:2006us,Berger:2008sj,Cascioli:2011va,Alwall:2014hca,Actis:2016mpe} and, more recently, to next-to-next-to-leading order (NNLO) in QCD.\footnote{
At least as far as processes with relatively  simple final states are concerned.}
While calculations at NNLO are typically sufficient to match the foreseeable precision of present and future LHC measurements, there is a handful of interesting processes for which theoretical predictions at even higher orders of perturbative QCD (i.e. N$^3$LO QCD) are warranted. 
This may happen for several reasons. 
Indeed, in some cases the convergence of the perturbative expansion in the strong coupling constant $\alpha_s$ turns out to be so slow that even NNLO QCD predictions have  a sizable uncertainty. 
Prominent examples of such a situation are processes where  color-singlet final states are produced in gluon fusion. 
For the important case of Higgs boson production in gluon fusion, it was explicitly shown that N$^3$LO QCD corrections are crucial to stabilize  theoretical uncertainties at the few percent level~\cite{Anastasiou:2015ema}.
In other cases, e.g. the Drell-Yan process, large statistics and clean final-state signatures led to experimental measurements with very high precision that is posed to increase further during Run III and the high-luminosity phase of the LHC.  
A theoretical description of the Drell-Yan process with matching or better precision remains a formidable challenge for the theory community.

The theoretical efforts aimed at extending the current computational technology to enable it to handle N$^3$LO calculations have recently culminated in the computation of the N$^3$LO QCD corrections to Higgs boson production in gluon fusion at the LHC~\cite{Anastasiou:2015ema,Anastasiou:2016cez,Mistlberger:2018etf,Dulat:2018bfe}.
Since these computations deal with a relatively simple final state and aim at calculating inclusive quantities, it is possible to employ the method of reverse unitarity~\cite{Anastasiou:2002yz} to simplify them.\footnote{Recently, an approximated N$^3$LO differential calculation for Higgs production has been completed using the $q_T$-subtraction formalism~\cite{Cieri:2018oms}.} 
Although the calculation of the N$^3$LO QCD corrections to the Higgs boson production cross section is a landmark in  perturbative computations in collider physics, the extension of the methods used in that computation to more complicated final states and more differential observables does not appear to be straightforward and it is interesting to think about alternative options. 

For definiteness, let us consider the production of a color-singlet final state $V$ in proton-proton collisions $pp \to V$.
Quite generally, the description of this process at N$^3$LO in QCD requires the knowledge of the NNLO QCD corrections to the production of  $V$ together with an additional QCD jet, $pp \to V + j$. 
The difference between $pp \to V+j$ at NNLO QCD and $pp \to V$ at N$^3$LO QCD is that the jet in the former case can become unresolved and that the virtual corrections to $pp \to V$ have no counterpart in the $pp \to V+j$ calculation. 
Since the difference between the two calculations appears in the kinematic regions where the color-singlet final state barely recoils against the QCD radiation, one can imagine partitioning the phase space into regions with and without recoil, using the NNLO QCD prediction for $pp \to V+j$ in the former region and studying the virtual corrections together with soft and collinear QCD radiation in the latter. 
This is the essence of a so-called \emph{slicing method}.   
For colorless final states, a widely used variable to slice the phase space into resolved and unresolved regions is the transverse momentum of the color-singlet $V$~\cite{Catani:2007vq}. 
More recently, the so-called $N$-jettiness observable~\cite{Stewart:2010tn,Gaunt:2015pea,Boughezal:2015dva} has allowed to generalize this idea to cases with final-state jets.
In the current paper we will focus on the latter variable and, in particular, on the case of $0$-jettiness, which is required to describe the inclusive production of a color-singlet final state.

To this end, we consider the process $pp \to V + X$, where $X$ represents the final-state QCD radiation. 
We denote the momenta of the incoming and outgoing partons by $p_{1,2}$ and $k_{1,..,n}$, respectively, and write the $0$-jettiness variable as 
\begin{equation}
\mathcal{T} = \sum\limits_{j=1}^{n} {\rm min}_{i \in \{1,2\}} \left[ \frac{2 p_i \cdot k_j}{Q_i}\right]\,. \label{eq:zerojet}
\end{equation}
In \cref{eq:zerojet},  $Q_{1,2}$ are the so-called hardness variables for the initial-state partons; they can be chosen in different ways and they are not relevant for the following discussion. 
The $0$-jettiness variable $\mathcal{T}$ has two important properties that allow one to use it as a slicing variable.
Indeed, it follows from the definition \cref{eq:zerojet} that $\mathcal{T} = 0$ in the absence of resolved  QCD radiation, i.e. for the process $pp \to V$. 
However, in the presence of any resolved QCD radiation one finds that $\mathcal{T} > 0$.
We can therefore introduce a cut-off $\mathcal{T}_0$ and divide the phase space for $V+X$ into two disjoint parts. 
We write schematically 
\begin{equation}
\sigma_{pp \to V+X}^{\rm N^3LO} = \sigma_{pp \to V+X}^{\rm N^3LO}\left( \mathcal{T} \leq \mathcal{T}_0 \right) 
+ \sigma_{pp \to V+X}^{\rm NNLO}\left( \mathcal{T} > \mathcal{T}_0 \right)\,. \label{eq:sigmasplit}
\end{equation}
Note the $\rm NNLO$ subscript in the second term on the right-hand side in \cref{eq:sigmasplit}; the reason for its appearance is that by imposing the $\mathcal{T} > \mathcal{T}_0$ constraint, we exclude the situation where {\it all} final-state partons become unresolved so that the calculation for $\mathcal{T} > \mathcal{T}_0$ reduces to the computation of the NNLO QCD corrections to $pp \to V+j$. 
Such calculations have already been performed for a variety of final states and we consider them to be known~\cite{Boughezal:2015dra,Ridder:2015dxa,Boughezal:2015aha,Boughezal:2015dva,Chen:2016zka,Gehrmann-DeRidder:2017mvr}.

On the other hand, the first term on the right-hand side of \cref{eq:sigmasplit} still receives contributions from those regions of phase space where the final-state radiation is fully unresolved.
In general, the computation of these contributions can be as difficult as the full N${}^3$LO calculation itself. 
However, for 0-jettiness, this does not happen. 
Indeed, it was shown in Ref.~\cite{Stewart:2010tn} that the cross section for $pp \to V+X$ simplifies substantially in the limit $\mathcal{T} \to 0$ and can be written as a convolution of the hard cross section for $pp \to V$ with the so-called beam and soft functions~\cite{Berger:2010xi,Gaunt:2014xga,Gaunt:2014cfa}.
The cross section reads
\begin{equation}
\lim_{\mathcal{T}_0 \to 0} 
d\sigma_{pp \to V+X}^{\rm N^3LO}\left( \mathcal{T} \leq \mathcal{T}_0 \right) \sim \, B \otimes B \otimes S \otimes d\sigma_{pp \to V}^{\rm N^3LO}\,,
\label{eq:fact0jet}
\end{equation}
where the two functions $B$ stand for the beam functions associated with each of the initial-state partons and $S$ represents the soft function. The general factorization formula for 
$N$-jettiness was 
originally derived in SCET~\cite{Bauer:2000ew,Bauer:2000yr,Bauer:2001ct,Bauer:2001yt,Bauer:2002nz}. The factorization of soft and collinear radiation, made apparent in \cref{eq:fact0jet}, is the key property of the 0-jettiness variable that simplifies the calculation of the differential cross section in the small-$\mathcal{T}$ limit. 

The cross-section formula \cref{eq:fact0jet} implies that, in order to employ the $0$-jettiness slicing to compute the N$^3$LO corrections to $pp \to V+X$, the beam and soft functions must be known at the same perturbative order. 
While the soft function is a purely perturbative object and can, at least in principle, be computed order-by-order in perturbation theory, the beam-function computation requires a convolution of perturbative \emph{matching coefficients} $I_{ij}$ with the non-perturbative parton distribution functions (pdfs) $f_{j}$ 
\begin{equation}
B_i = \sum_{\rm partons~j}\, I_{ij} \otimes f_j\,, \quad \mbox{where} \quad i,j=\{q,\bar{q},g\} \label{eq:matchI}.
\end{equation}

The computation of the N$^3$LO QCD corrections to the matching coefficient $I_{qq}$ is the main topic of this paper.
At three loops, $I_{qq}$ receives contributions from three classes of partonic subprocesses: the emission of three collinear partons, which we will refer to as the triple-real contribution (RRR); the one-loop corrections to the emission of two collinear partons, or the double-real single-virtual contribution (RRV); and, finally, the two-loop virtual corrections to the emission of one collinear parton, or the single-real double-virtual contribution (RVV).

In a previous paper~\cite{Melnikov:2018jxb}, we presented the master integrals required for the calculation of the RRV contribution with two emitted gluons to the matching coefficient $I_{qq}$. 
In this paper, we focus on the master integrals required for the computation of the RRR contribution to the matching coefficient that originate from the process where the initial-state quark emits three collinear gluons before entering the hard scattering process.
We note that the same master integrals can be used to compute the $N_f$-enhanced triple-real contribution to $I_{qq}$, caused by the emission of a gluon and a quark-antiquark pair collinear to the initial-state quark.

The rest of the paper is organized as follows: in \cref{sec:amp} we explain how to compute the RRR contribution to the matching coefficient $I_{qq}$ by considering collinear limits of scattering amplitudes and how reverse unitarity can be used to reduce this calculation to the computation of a large set of three-loop master integrals.
We then show in \cref{sec:masters} how these integrals can be computed using the method of differential equations. 
In \cref{sec:checks}, we explain how the calculation was validated and we present our final results in \cref{sec:results}. 
We conclude in \cref{sec:concl}.
The list of master integrals can be found in \cref{app:listofmasters}.
Some peculiar identities among master integrals are described in \cref{app:extraids}.
The results for the master integrals are provided in computer-readable format in an ancillary file, which is available at \url{https://www.ttp.kit.edu/_media/progdata/2019/ttp19-009.tar.gz}.


\section{Matching coefficient}
\label{sec:amp} 

In this section we discuss how to compute the N${}^3$LO contributions to  the matching coefficient $I_{qq}$ for the 0-jettiness beam function. 
Since the matching coefficients describe the physics of collinear emissions off the incoming partons, they can be calculated by integrating the collinear limits of the corresponding scattering amplitudes squared, over the phase space restricted by the fixed value of the 0-jettiness variable. 

More specifically, the phase-space integration must be performed by imposing constraints on the transverse virtuality of the collinear partons and on the light-cone momentum of the parton that enters the hard-scattering process~\cite{Stewart:2010tn}.
Since singular collinear emissions factorize on the external lines, the hard-scattering process decouples. 
The collinear emissions are described by splitting functions; for this reason, the relevant contributions to the matching coefficients can be computed by integrating these functions over a restricted phase space~\cite{Ritzmann:2014mka}.
This observation is particularly useful since the prescription for computing the splitting functions to any order in the strong coupling constants has been laid out in Ref.~\cite{Catani:1999ss}.

\begin{figure}
\begin{subfigure}{0.50\textwidth}
\includegraphics[width=\linewidth]{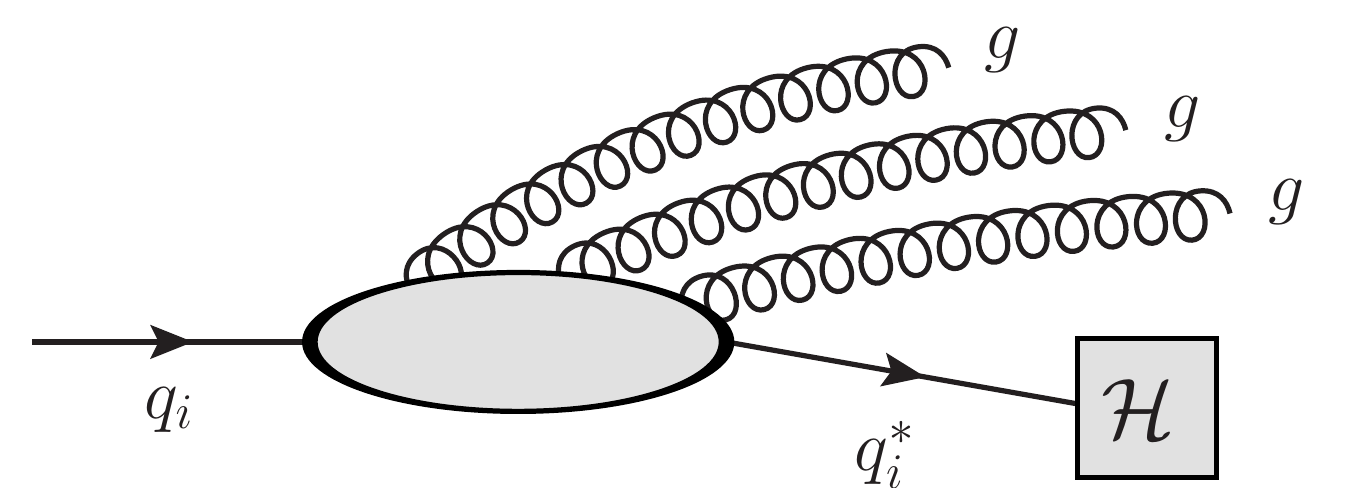}
\caption{}
\label{fig:proc_ggg}
\end{subfigure}
\hspace*{\fill}
\begin{subfigure}{0.48\textwidth}
\includegraphics[width=\linewidth]{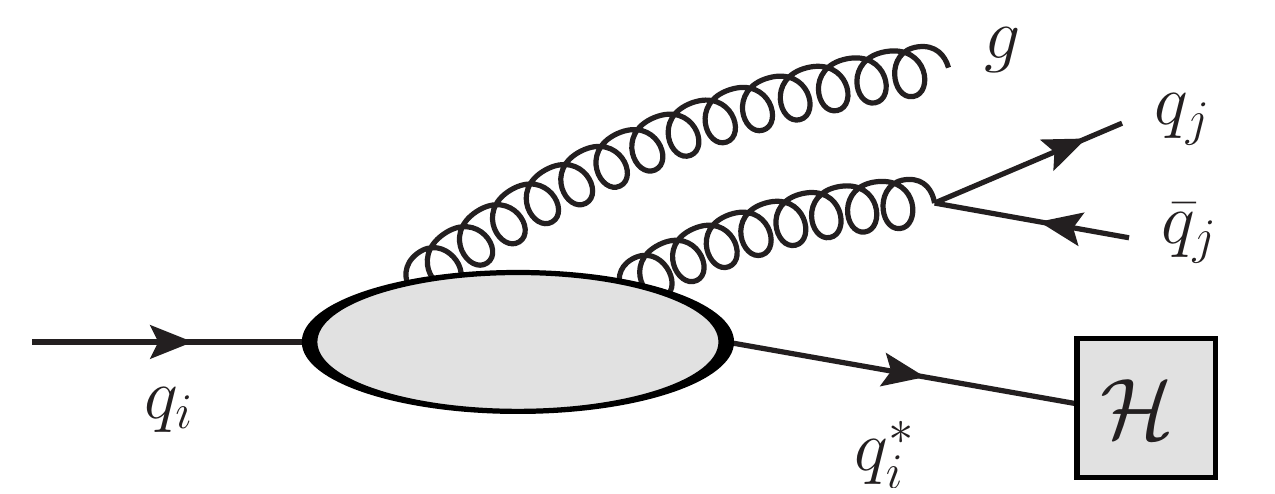}
\caption{} 
\label{fig:proc_qqbg_Nf}
\end{subfigure}
\vspace{-5mm}
\caption{The process $q_i \to q_i^* + ggg $ and the process $q_i \to q_i^* + q_j \bar{q}_j g$ for $i \ne j$.}
\label{fig:processes}
\end{figure}

Focusing on the triple-real (RRR) contribution to the matching coefficient $I_{qq}$, we need to consider the tree-level splitting of a quark into a virtual quark of the same flavor and three collinear partons. 
These three partons can be either three gluons or a quark-antiquark pair and a gluon, so that there are two generic possibilities: $q_i \to q_i^* + ggg $ and $q_i \to q_i^*+ q_j \bar{q}_j + g$.
In this paper we consider the process involving three collinear gluons as well as the process involving a collinear gluon and a collinear quark-antiquark pair of a different flavor with respect to the incoming quark, i.e. $i \neq j$, see \cref{fig:processes}.
The case $i=j$ requires additional contributions that are not considered in this paper.
However, it is easy to see that the neglected contributions are sub-leading in the $N_c \to \infty$ limit, where $N_c$ is the number of colors, and in the $N_f \to \infty$ limit, where $N_f$ is the number of massless quark flavors.
Hence, even neglecting the $i=j$ contributions, we can obtain the result for $I_{qq}$ that is valid in the large-$N_c$ or large-$N_f$ limits.
In the remainder of this section, we focus our discussion on the process in \cref{fig:proc_ggg} for definiteness.

We can now describe the details of the calculation. 
We follow the discussion in Ref.~\cite{Melnikov:2018jxb}, where the master integrals for the double-real single-virtual contribution to $I_{qq}$ were computed.
We consider a massless quark with momentum $p$ which emits three collinear gluons with momenta $k_i$, $i=1,2,3$, and  enters the hard process with momentum $p^*$
\begin{equation}
q(p) \to q^*(p^*) + g(k_1) + g(k_2) + g(k_3)\,, \qquad  p^* = p -k_1 -k_2 -k_3\,.
\end{equation}
As we already explained, the relevant contribution to the matching coefficient is obtained by integrating the $q \to q^* + ggg$ splitting function over the phase space of the emitted gluons with appropriate constraints.
In order to write these constraints in a convenient form, we fix the component of the momentum $p^*$ along the momentum of the incoming quark $p$ and write
\begin{equation}
p^{*\mu} = z p^\mu + y \bar{p}^\mu + k_\perp^\mu\,, \quad k_{123}^\mu = (1-z) p^\mu - y \bar{p}^\mu - k_\perp^\mu\,. 
\label{eq:colllim}
\end{equation}
In \cref{eq:colllim}, we used $k_{123} = \sum\limits_{i=1}^{3} k_i^\mu$.
We also introduced a light-cone momentum $\bar{p}$, which is complementary to $p$ so that $\bar{p}^2 = 0$ and $p \cdot k_\perp = \bar{p} \cdot k_\perp = 0$. 
The emitted gluons are on the mass shell, i.e. $k_i^2 = 0$ for $i=1,2,3$. 
With these definitions we have $y = - (p \cdot k_{123})/(p \cdot \bar{p})$.

We now introduce the transverse virtuality $t = -((p^*)^2 - k_\perp^2)$ and, using the above results, write it as
\begin{equation}
t = -z y \,2 p\cdot \bar{p} = z \, 2 p \cdot k_{123}\,. 
\label{eq:constr1}
\end{equation}
Note that,  in the case of collinear emissions, $t \sim \mathcal{T}$.
We also impose a constraint on the light-cone component of the momentum of the quark that enters the hard process.
We write it as
\begin{equation}
s (1-z) = 2 \bar{p} \cdot k_{123}\,, \quad \mbox{with} \quad s = 2 p \cdot \bar{p}\,.
\label{eq:constr2}
\end{equation}
Using \cref{eq:constr1,eq:constr2}, we write the generic contribution of the three-gluon final state to the matching coefficient $I_{qq}$ in the following way
\begin{align}
I_{qq}(t,s,z) \sim  \int &\prod_{i=1}^3 \left[ \frac{d^d k_i}{(2\pi)^{d-1} } \delta^+(k_i^2) \right]  \delta\left( p\cdot k_{123} - \frac{t}{2z} \right)   
\delta\left( \bar{p}\cdot k_{123} - \frac{s(1-z)}{2} \right)  \nonumber \\
&\times P_{qq}(p,\bar{p},\{k_i\})\,.
\label{eq:Iqq}
\end{align}
In \cref{eq:Iqq} the integrand $P_{qq}(p,\bar{p},\{k_i\})$ describes the $q \to q^* + ggg$ splitting function. 
Below we explain how to compute it.

\begin{figure}
\begin{subfigure}{0.48\textwidth}
\includegraphics[width=\linewidth]{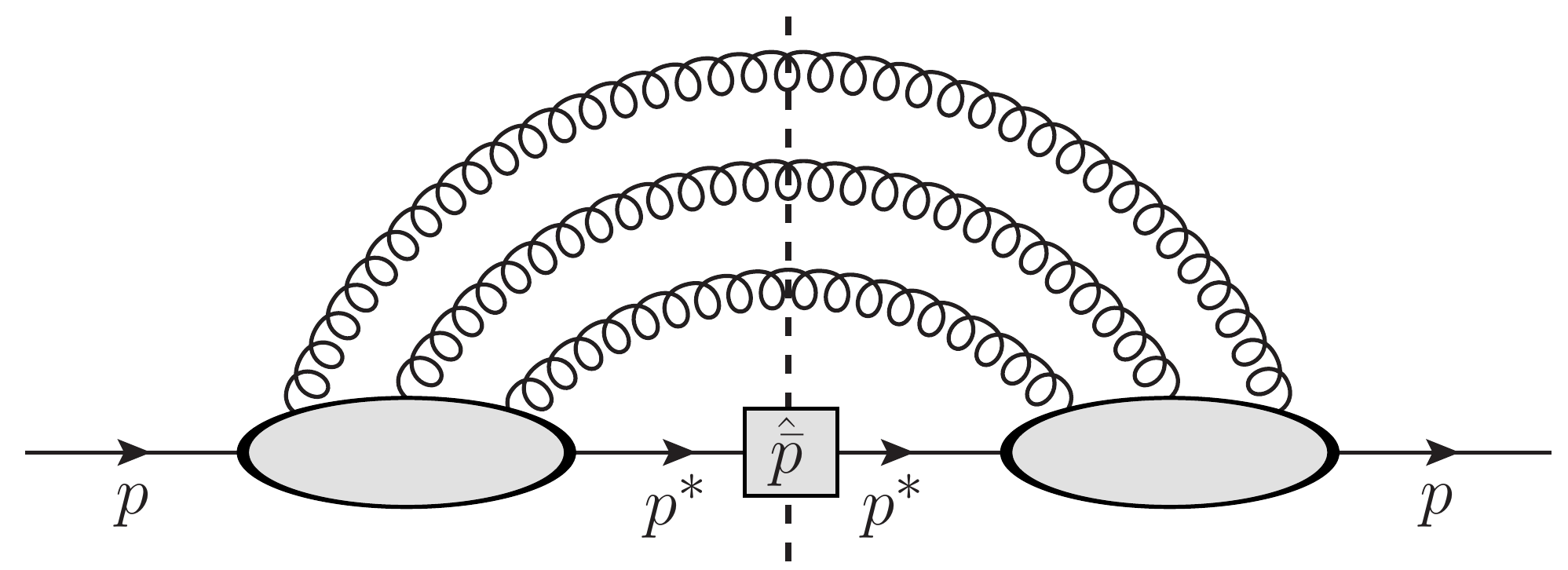}
\caption{}
\label{fig:amp_ggg}
\end{subfigure}
\hspace*{\fill}
\begin{subfigure}{0.48\textwidth}
\includegraphics[width=\linewidth]{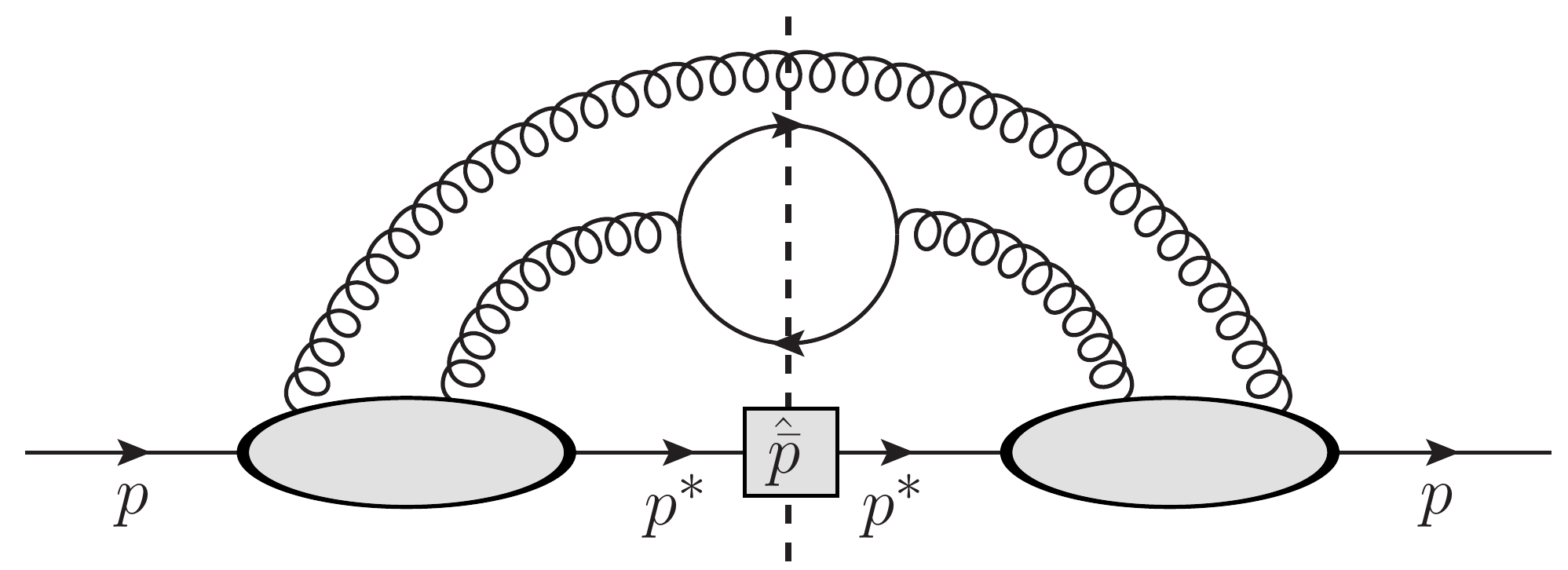}
\caption{}
\label{fig:amp_qqbg}
\end{subfigure}
\vspace{-5mm}
\caption{The collinear projection of the squared scattering amplitude for the process $q \to q^* + ggg $ and the process $q \to q^* + q' \bar{q}' g$ for $q' \ne q$.} 
\label{fig:amplitudes}
\end{figure}

As described in Ref.~\cite{Catani:1999ss}, the $q \to q^* + ggg$ splitting function can be obtained as the collinear projection of the squared scattering amplitude for the corresponding process \cref{fig:proc_ggg}.
To this end, we generate the scattering amplitude as a sum of all diagrams that contribute to the $q \to q^* + ggg$ process. 
The diagrams are turned into mathematical expressions with the standard QCD Feynman rules, albeit with a symbolic placeholder for the arbitrary hard-scattering process.
The axial gauge is chosen for the gluons, both internal and external ones, and the light-cone vector $\bar{p}$ from \cref{eq:colllim} is selected as the corresponding gauge-fixing vector.
Squaring the amplitude, we produce a Dirac trace of the form $\mathrm{Tr}[\dotsm \, \hat{p}^* \, \hat{\mathcal{H}} \, \hat{p}^* \, \dotsm ]$, where $\hat{p}^* = \gamma^\mu p^*_\mu$ and $p^*$ is the momentum that enters the hard scattering process. 
The Dirac matrix $\hat{\mathcal{H}}$ is a symbolic representation for the (product of) gamma matrices in the hard interaction.
The collinear projection of the squared scattering amplitude, schematically depicted in \cref{fig:amplitudes}, is achieved by making the replacement
\begin{align}
\mathrm{Tr}[\dotsm \, \hat{p}^* \, \hat{\mathcal{H}} \, \hat{p}^* \, \dotsm ] 
\rightarrow
\mathrm{Tr}[\dotsm \, \hat{p}^* \, \hat{\bar{p}} \, \hat{p}^* \, \dotsm ] \,,
\label{eq:collinear_projection}
\end{align}
which has the effect of removing all non-singular contributions in the limit where all three gluons become collinear to the incoming quark.

In practice, we generate the diagrams that contribute to the process $q(p) \to q^*(p^*) + g(k_1) + g(k_2) + g(k_3)$ with QGRAF~\cite{Nogueira:1991ex}.
We perform the relevant Dirac and Lorentz algebra in FORM~\cite{Vermaseren:2000nd} and Mathematica in two independent implementations.
Since we work in the axial gauge with the gauge-fixing vector $\bar{p}$, the sum over polarizations for a gluon with momentum $k_i$ reads 
\begin{equation}
\sum_{\rm pol} \epsilon_i^{\mu}(k_i) \left(  \epsilon_i^{\nu}(k_i) \right)^* = - g^{\mu \nu} + \frac{k_i^\mu \bar{p}^\nu + k_i^{\nu} \bar{p}^\mu}{k_i \cdot \bar{p}}\,,\quad
\mbox{for}\;\; i=1,2,3\,. \label{eq:axialgauge}
\end{equation}
After applying the collinear projection in \cref{eq:collinear_projection}, the squared amplitude can be written as a linear combination of a large number of scalar phase-space integrals of the following form
\begin{align}
\mathcal{I} = \int \prod_{i=1}^3 \left[ \frac{d^d k_i}{(2\pi)^{d-1} } \delta^+(k_i^2) \right] \frac{ \delta\left( p\cdot k_{123} - \frac{t}{2z} \right)   
\delta\left( \bar{p}\cdot k_{123} - \frac{s(1-z)}{2} \right)    \mathcal{N}}{D_1^{n_1} \cdots D_t^{n_t}}\,. \label{eq:genI}
\end{align}
Here, $\mathcal{N}$ is a generic combination of scalar products of the parton momenta, and $D_j$ are propagators, including linear propagators that originate e.g. from the denominators in \cref{eq:axialgauge}.
These integrals can be computed efficiently using the method of reverse unitarity~\cite{Anastasiou:2002yz}, which allows one to turn the delta function constraints in \cref{eq:genI} into cut propagators, mapping the problem of computing phase-space integrals onto the calculation of a large number of three-loop Feynman integrals.

We need to organize these integrals into \emph{integral families} to enable the reduction to master integrals through the integration-by-parts identities (IBPs)~\cite{Tkachov:1981wb,Chetyrkin:1981qh,Laporta:2001dd}.
As is often the case when dealing with phase-space integrals in the framework of reverse unitarity, this step is not entirely straightforward. 
Indeed, a well-defined integral family requires as many propagators as the number of independent scalar products in the problem at hand.
In our case there are two independent external momenta $p$ and $\bar{p}$ and three gluon momenta $k_i$. 
This implies that any integral family must contain exactly $12$ independent propagators.
By directly inspecting the Feynman diagrams, it is easy to see that, after accounting for the delta function from the $0$-jettiness constraint, many diagrams do generate scalar integrals of the form shown in \cref{eq:genI}, but with more than $12$ different propagators.

To remedy this problem, we need to use partial fractioning. 
For example, it may happen that an integral contains all three linear propagators $1/k_i\cdot \bar{p}$ with $i=1,2,3$.
However, the $0$-jettiness constraint in \cref{eq:genI} implies that the three propagators $1/k_i\cdot \bar{p}$ are not linearly independent.
Indeed, we can write
\begin{align}
\frac{1}{k_1 \cdot \bar{p}\,\, k_2 \cdot \bar{p}\,\, k_3 \cdot \bar{p}} = \frac{2}{1-z} 
\left[ \frac{1}{k_1 \cdot \bar{p}\, k_2 \cdot \bar{p}} 
+ \frac{1}{k_1 \cdot \bar{p}\, k_3 \cdot \bar{p}} 
+ \frac{1}{k_2 \cdot \bar{p}\, k_3 \cdot \bar{p}} \right]\,,
\end{align}
which allows us to reduce the number of propagators by one.

Unfortunately, this procedure is ambiguous, since different ways of partial fractioning can lead to different integral families and different integrals. 
While it is usually sufficient to use the IBP identities to remove most of this redundancy, some of the integrals that appear to be independent under IBPs can still be related by special partial fractioning identities and we need to separately account for that possibility.

Due to the ambiguity mentioned above, we find it convenient to introduce an overcomplete set of integral families in order to simplify the mapping of diagrams to topologies. 
Nevertheless, performing the IBP reduction and accounting for additional identities that originate from the partial fractioning, we find that all diagrams can be expressed in terms of $91$ master integrals which are drawn from 19 different topologies, see \cref{tab:inverseprops}.
We performed the reduction to master integrals using Reduze~\cite{vonManteuffel:2012np} and KIRA~\cite{Maierhoefer:2017hyi}, both of which support the generation and solution of IBPs for Feynman integrals with cut propagators, and we verified that the results of the two reduction codes are equivalent.

We use the following notation for the master integrals
\begin{align}
\mathcal{I}^{\rm top}_{n_1,n_2,n_3,n_4,n_5,n_6,n_7} = \int \mathfrak{D}^d k_1 \mathfrak{D}^d k_2 \mathfrak{D}^d k_3\, 
\frac{\delta\big(p \cdot k_{123} - \frac{t}{2z}\big) \delta\big(\bar{p} \cdot k_{123} - \frac{s(1-z)}{2}\big)}{D_1^{n_1} D_2^{n_2}D_3^{n_3}D_4^{n_4}D_5^{n_5}D_6^{n_6}D_7^{n_7} }\,,
\label{eq:Itop}
\end{align}
where $d=4-2\eps$ and the subscript `top' indicates one of the topologies in \cref{tab:inverseprops} where the inverse propagators $D_i$ for each topology are defined.
The integration measure for each final-state particle reads
\begin{equation}
\mathfrak{D}^d k_i =
\frac{d^d k_i}{(2 \pi)^{d-1}} \delta^{+}(k_i^2)\,.
\end{equation}
We use these notations to present the list of master integrals in \cref{app:listofmasters}.

\begin{table}[ht!]
\begin{tabular}[c]{l|lllllll}
\text{top} ~~&~
$D_1$ \hspace{13mm} & 
$D_2$ \hspace{13mm} & 
$D_3$ \hspace{13mm} & 
$D_4$ \hspace{15mm} & 
$D_5$ \hspace{15mm} & 
$D_{6}$ \hspace{6mm} & 
$D_{7}$ \hspace{6mm}  \\
\hline
 $A_1$ & ~~$(p-k_1)^2$ & $(p-k_2)^2$ & $(p-k_{12})$ & $(p-k_{13})^2$ & $(p-k_{123})^2$ & $\pbar\cdot k_1$ & $\pbar\cdot k_2 $\\
 $A_2$ & ~~$(p-k_1)^2$ & $(p-k_2)^2$ & $(p-k_{12})$ & $(p-k_{13})^2$ & $(p-k_{123})^2$ & $\pbar\cdot k_1$ & $\pbar\cdot k_3 $\\
 $A_3$ & ~~$(p-k_1)^2$ & $(p-k_2)^2$ & $(p-k_{12})$ & $(p-k_{13})^2$ & $(p-k_{123})^2$ & $\pbar\cdot k_2$ & $\pbar\cdot k_3 $\\
 $A_4$ & ~~$k_{12}^2$ & $(p-k_1)^2$ & $(p-k_{12})$ & $(p-k_{13})^2$ & $(p-k_{123})^2$ & $\pbar\cdot k_1$ & $\pbar\cdot k_2 $\\
$A_5$ & ~~$k_{12}^2$ & $(p-k_1)^2$ & $(p-k_{12})$ & $(p-k_{13})^2$ & $(p-k_{123})^2$ & $\pbar\cdot k_1$ & $\pbar\cdot k_3 $\\
$A_6$ & ~~$k_{12}^2$ & $(p-k_1)^2$ & $(p-k_{12})$ & $(p-k_{13})^2$ & $(p-k_{123})^2$ & $\pbar\cdot k_2$ & $\pbar\cdot k_3 $\\
 $A_7$ & ~~$(p-k_1)^2$ & $(p-k_2)^2$ & $(p-k_{13})$ & $(p-k_{23})^2$ & $(p-k_{123})^2$ & $\pbar\cdot k_1$ & $\pbar\cdot k_2 $\\
$ A_8$ & ~~$k_{12}^2$ & $k_{13}^2$ & $(p-k_2)^2$ & $(p-k_{23})^2$ & $(p-k_{123})^2$ & $\pbar\cdot k_1$ & $\pbar\cdot k_{2} $\\
 $A_9$ & ~~$k_{12}^2$ & $k_{13}^2$ & $(p-k_2)^2$ & $(p-k_{23})^2$ & $(p-k_{123})^2$ & $\pbar\cdot k_1$ & $\pbar\cdot k_{3} $\\
 $A_{10}$ & ~~$k_{12}^2$ & $(p-k_1)^2$ & $(p-k_{3})^2$ & $(p-k_{13})^2$ & $(p-k_{123})^2$ & $\pbar\cdot k_1$ & $\pbar\cdot k_2 $\\
 $A_{11}$ & ~~$k_{12}^2$ & $(p-k_1)^2$ & $(p-k_3)^2$ & $(p-k_{13})^2$ & $(p-k_{123})^2$ & $\pbar\cdot k_2$ & $\pbar\cdot k_3 $\\
 $A_{12}$ & ~~$k_{12}^2$ & $k_{13}^2$ & $k_{123}^2$ & $(p-k_{2})^2$ & $(p-k_{12})^2$ & $\pbar\cdot k_1$ & $\pbar\cdot k_2 $\\
 $A_{13}$ & ~~$k_{12}^2$ & $k_{13}^2$ & $k_{123}^2$ & $(p-k_{2})^2$ & $(p-k_{12})^2$ & $\pbar\cdot k_1$ & $\pbar\cdot k_3 $\\
 $A_{14}$ & ~~$(p-k_1)^2$ & $(p-k_2)^2$ & $(p-k_{12})^2$ & $(p-k_{13})^2$ & $(p-k_{123})^2$ & $\pbar\cdot k_1$ & $\pbar\cdot k_{13} $\\
 $A_{15}$ & ~~$k_{12}^2$ & $(p-k_1)^2$ & $(p-k_{12})^2$ & $(p-k_{13})^2$ & $(p-k_{123})^2$ & $\pbar\cdot k_1$ & $\pbar\cdot k_{12} $\\
 $A_{16}$ & ~~$k_{12}^2$ & $k_{13}^2$ & $(p-k_{2})^2$ & $(p-k_{23})^2$ & $(p-k_{123})^2$ & $\pbar\cdot k_1$ & $\pbar\cdot k_{12} $\\
 $A_{17}$ & ~~$k_{12}^2$ & $k_{13}^2$ & $(p-k_2)^2$ & $(p-k_{23})^2$ & $(p-k_{123})^2$ & $\pbar\cdot k_1$ & $\pbar\cdot k_{13} $\\
 $A_{18}$ & ~~$k_{12}^2$ & $k_{13}^2$ & $(p-k_{12})^2$ & $(p-k_{13})^2$ & $(p-k_{123})^2$ & $\pbar\cdot k_1$ & $\pbar\cdot k_{12} $\\
 $A_{19}$ & ~~$k_{12}^2$ & $k_{13}^2$ & $k_{123}^2$ & $(p-k_{2})^2$ & $(p-k_{12})^2$ & $\pbar\cdot k_{13}$ & $\pbar\cdot k_{3} $
\end{tabular}
\caption{The inverse propagators $D_i$ for each of the 19 topologies $A_1 \dotsc A_{19}$. Here we use the shorthand notation $k_{ij} = k_i + k_j$ and $k_{ij\ell} = k_i + k_j + k_\ell$.}
\label{tab:inverseprops}
\end{table}

While the set of master integrals shown in \cref{eq:listofmasters} is indeed minimal with respect to the IBPs, we were able to find two additional relations between them, that do not follow from IBPs and partial fractioning. 
These identities read
\begin{align}
\mathcal{I}^{A_{10}}_{1,1,1,0,1,0,1} + \frac{1}{2} \frac{(1-z)}{z}\;  \mathcal{I}^{A_{6}}_{1,1,0,0,1,1,1} = 0\,, 
\label{eq:extraidentity1}\\
\mathcal{I}^{A_{1}}_{0,0,1,0,0,1,0} + \frac{2(1-2\epsilon)}{\epsilon} \frac{1}{z}\;  \mathcal{I}^{A_{12}}_{0,0,1,0,1,0,0} = 0\,.
\label{eq:extraidentity2}
\end{align}
They allow us to reduce the number of independent master integrals from $91$ to $89$.
Nevertheless, we prefer to compute the full set of $91$ master integrals and verify these identities a posteriori. 
We note that these identities can be proven by studying the differential equations satisfied by the four master integrals that appear in \cref{eq:extraidentity1,eq:extraidentity2} together with the direct inspection of their integral representations. 
We describe the proof in \cref{app:extraids}.


\section{Master integrals}
\label{sec:masters}

The master integrals defined in \cref{eq:Itop} depend on $t$, $z$ and $s = 2 p \cdot \pbar$. 
However, the dependence on $s$ and $t$ is trivial.
This becomes manifest after the simultaneous re-scaling $k_i \to k_i \, \sqrt{t}$, $p \to p \, \sqrt{t}$ and $\pbar \to \pbar \, s / \sqrt{t}$.
The re-scaling has the effect of extracting  powers of $s$ and $t$ from each integral, leaving only a non-trivial dependence
on $z$.
Explicitly, we find
\begin{align}
\begin{aligned}
\mathcal{I}^{\rm top}_{n_1,n_2,n_3,n_4,n_5,n_6,n_7} 
&= \int \mathfrak{D}^d k_1 \mathfrak{D}^d k_2 \mathfrak{D}^d k_3\, 
\frac{\delta \big(p \cdot k_{123} - \frac{t}{2z} \big)\delta \big(\bar{p} \cdot k_{123} - \frac{s(1-z)}{2} \big)}{D_1^{n_1} D_2^{n_2}D_3^{n_3}D_4^{n_4}D_5^{n_5}D_6^{n_6}D_7^{n_7} }
\\
&= 
s^{-\mathcal{M}} \, t^{\mathcal{N}}
\int \mathfrak{D}^d k_1 \mathfrak{D}^d k_2 \mathfrak{D}^d k_3\, 
\frac{\delta\big(p \cdot k_{123} - \frac{1}{2} \big)\delta \big(\bar{p} \cdot k_{123} - \frac{(1-z)}{2} \big)}{D_1^{n_1} D_2^{n_2}D_3^{n_3}D_4^{n_4}D_5^{n_5}D_6^{n_6}D_7^{n_7} }\,
\,,
\label{eq:after_scaling_st}
\end{aligned}
\end{align}
where $\mathcal{M} = 1+n_6+n_7$ and $\mathcal{N} = 5-3\eps-\sum\limits_{i=1}^{5} n_i$.
As a consequence, we can set $s = t = 1$ everywhere and focus only on the $z$-dependence of the master integrals.

We determine the $z$-dependence of the master integrals with the method of differential equations~\cite{Kotikov:1990kg,Remiddi:1997ny,Gehrmann:1999as}.
To this end, we differentiate each of the master integrals with respect to $z$ and express the result in terms of master integrals using integration-by-parts identities.
We collect the master integrals into a vector $\vec{\mathcal{I}}(z,\eps)$ and write the resulting closed system of differential equations as
\begin{align}
\frac{d}{dz} \vec{\mathcal{I}}(z,\eps) = \hat A(z,\eps)\, \vec{\mathcal{I}}(z,\eps)\,.
\label{eq:DEQv1}
\end{align}
The entries of the matrix $\hat A(z,\eps)$ are rational functions of $z$ and $\eps$.

The complexity of these differential equations depends strongly on the explicit form of the matrix $\hat A(z,\eps)$, which, in turn, depends on the choice of the master integrals.
Our goal is to choose the master integrals in such a way that the matrix becomes canonical and Fuchsian~\cite{Kotikov:2012ac,Henn:2013pwa,Lee:2014ioa}, $\hat A(z,\eps) = \eps \sum\limits_{z_0} \frac{\hat A_{z_0}}{z-z_0}$. 
Note that the matrices $\hat A_{z_0}$ should be both $z$- and $\epsilon$-independent. 
If such a form is found, the process of solving differential equations simplifies greatly. 

It turns out, however, that the system in \cref{eq:DEQv1} cannot be brought to a canonical Fuchsian form without replacing $z$ with a more suitable variable.
Indeed, it is easy to see that upon integration, the homogeneous terms of some of the differential equations produce the square root $\sqrt{z(4-z)}$. 
The presence of square roots complicates substantially the problem of finding a canonical Fuchsian form.
To rationalize it, we change variables from $z$ to $x$ according to the following equation
\begin{align}
z = \frac{(1+x)^2}{x}\,.
\label{eq:LandauVariable}
\end{align}
Having removed all square roots, we can construct the appropriate transformation $\vec{\mathcal{I}}(x,\eps) = \hat T(x,\eps)\,\vec{\mathcal{I}}_{\text{can}}(x,\eps)$ with the program \texttt{Fuchsia} \cite{Gituliar:2017vzm}.
As a result, we find
\begin{align}
\frac{d}{d x} \vec{\mathcal{I}}_{\text{can}}(x,\eps) = \eps \,
\left(\sum_{x_0}\frac{\hat A_{x_0}}{x-x_0}\right)
\vec{\mathcal{I}}_{\text{can}}(x,\eps)\,.
\label{eq:DEQv2}
\end{align}
The differential equations have singularities drawn from the list 
$x_0 \in \{-1, 0, 1, R_1^{\pm}, R_2^{\pm}, R_3^{\pm}\}$, which in turn correspond to singularities in $z$ given by $z_0 \in \{ 0, \infty, 4, 1, -1, 2 \}$.
The symbols $R_1^{\pm}$, $R_2^{\pm}$ and $R_3^{\pm}$ represent the two roots of each of the quadratic polynomials $P_1 = 1+x+x^2$, $P_2 = 1+3x+x^2$ and $P_3 = 1+x^2$, respectively.

It is convenient to solve  the system of differential equations \cref{eq:DEQv2} expanding  around $\eps = 0$.
We write  $\vec{\mathcal{I}}_{\text{can}}(x,\eps)$ as 
\begin{align}
\vec{\mathcal{I}}_{\text{can}}(x,\eps) = \hat M(x,\eps)\, \vec{B}(\eps)\,,
\label{eq:sol}
\end{align}
where $\vec{B}(\eps)$ are the integration constants. 
The $x$-dependence resides solely in the matrix $\hat M(x,\eps)$, whose elements have the form
\begin{align}
M_{ij}(x,\eps) = \sum_{k \geq 0} \, \sum_{\vec{w} \,\in\, W(k)} c_{i,j,k,\vec{w}} \,\, \eps^k \, G(\vec{w}; x)\,.
\label{eq:Mij}
\end{align}
We calculate the sum over $k$ up to and including $k=6$, corresponding to ${\mathcal O}( \eps^6)$, which is the highest order that will contribute to the finite part of the matching coefficient in the $\eps \to 0$ limit. 
For a given $k$, the inner sum in \cref{eq:Mij} runs over $W(k)$, containing all vectors $\vec w$ of the length $k$ with components drawn from the set of roots $\{-1, 0, 1, R_1^\pm, R_2^\pm, R_3^\pm\}$.
The functions $G(\vec{w}; x)$ are the Goncharov polylogarithms~\cite{Kummer,Goncharov:1998kja,Remiddi:1999ew,Goncharov:2001iea}
\begin{align}
G(w_1,w_2,\dotsc,w_n; x) = \int\limits_0^x  dt \,\frac{G(w_2,w_3,\dotsc,w_n ; t)}{t-w_1}\,.
\label{eq:gpl}
\end{align}
They can be evaluated numerically with the help of the program \texttt{Ginac} \cite{Vollinga:2004sn}.
Apart from the technical difficulty in handling large expressions, 
the construction of the matrix $\hat M(x,\eps)$ can be done in a relatively straightforward way.

On the contrary, the determination of the integration constants $\vec{B}(\eps)$ in \cref{eq:sol} is much less straightforward. 
We obtain them by analyzing the master integrals in the limit $z \to 1$. 
To this end, it is important to recognize that the master integrals significantly simplify in that limit. 
In particular, to leading order in $(1-z)$ we can replace the propagators  $1/(p-k_{ij})^2$ with\; 
$1/(-2 k_{ij} \cdot p)$.  
Note that this replacement renders the integrals uniform functions of the momenta $k_i$ so that, in the soft limit, the integral factorizes into a constant and a $(1-z)$-dependent factor.

The possibility to neglect $k_{ij}^2$ relative to $k_{ij} \cdot p$ follows from the following argument.
Let us select a frame in which the external momenta are $p = \frac{1}{\sqrt{2}}(1,0,0,1)$ and $\pbar = \frac{1}{\sqrt{2}}(1,0,0,-1)$ and introduce a Sudakov decomposition of the gluon momentum $k_i$
\begin{align}
k_i^\mu = \alpha_i \, p^\mu + \beta_i \, \pbar^\mu + k_{i \perp}^\mu\,.
\label{eq:Sudakov}
\end{align}
Since $\beta_i = k_i \cdot p = k_i^{0}p^{0} (1-\cos \theta_i)$ and  $\alpha_i = k_i \cdot \bar p = k_i^{0} \pbar^{0} ( 1+\cos \theta_i)$, we conclude that all $\alpha$'s and $\beta$'s are positive definite. 
According to the phase-space constraints \cref{eq:genI}, the sum $\alpha_{123} = \alpha_1+\alpha_2+\alpha_3$ goes to zero in the $z \to 1$ limit and, since all $\alpha$'s are positive, we conclude that each $\alpha_i$ goes to zero in that limit at least as fast as $\mathcal{O}(1-z)$.
In contrast, the sum of the $\beta_i$'s is constrained to be equal to one, so that up to two of them could vanish at $z = 1$.
We write
\begin{align}
\frac{1}{(p-k_{ij})^2} 
= \frac{1}{k_{ij}^2 - 2 k_{ij} \cdot p} \,,
\end{align}
where we have used that $p^2=0$.
In terms of the Sudakov parameters, $k_{ij}^2$ reads
\begin{align}
k_{ij}^2 = \alpha_i \beta_j + \alpha_j \beta_i - 2 \sqrt{\alpha_i \beta_j \alpha_j \beta_i} \cos \theta_{ij} 
\quad\text{and}\quad
2 k_{ij} \cdot p = \beta_i + \beta_j = \beta_{ij}\,,
\end{align}
where we have used $k_i^2 = k_j^2 = 0$.
Assuming that, in the limit $z \to 1$, each $\alpha_i = \mathcal{O}(1-z)$ and each $\beta_i = \mathcal{O}(1)$ we find $k_{ij}^2 = \mathcal{O}(1-z)$ and $2 k_{ij} \cdot p = \mathcal{O}(1)$.
Hence, we can neglect $k_{ij}^2$ relative to $2 k_{ij} \cdot p$.
The situation does not change, should any of the $\alpha_i$'s vanish faster than $\mathcal{O}(1-z)$.
Another possibility is that both $\beta_i$ and $\beta_j$ vanish as $\mathcal{O}(1-z)$, such that $2 k_{ij} \cdot p \to 0$. 
However, in that situation $k_{ij}^2$ scales as $\mathcal{O}((1-z)^2)$ or faster, and we can again neglect it relative to $2 k_{ij} \cdot p$.
Therefore, a replacement 
\begin{align}
\frac{1}{(p-k_{ij})^2}
\to 
\frac{1}{- 2 k_{ij} \cdot p} \,,
\label{eq:27km}
\end{align}
is valid in the $z \to 1 $ limit, to leading power in $(1-z)$. 

Since the replacement in \cref{eq:27km} implies that {\it all} propagators become uniform functions of 
the gluon momenta in the soft limit, the extraction of the $(1-z)$-dependence of any integral becomes straightforward. 
We note that, in  that limit, the phase-space constraints from \cref{eq:after_scaling_st} become $\delta\big(k_{123}\cdot p - \tfrac{1}{2}\big) \delta\big(k_{123}\cdot\pbar - \tfrac{(1-z)}{2}\big)$ and, upon re-scaling the momenta as $k_i \to k_i \sqrt{1-z} $, $\pbar \to \pbar \sqrt{1-z}$ and $p \to p/\sqrt{1-z}$, we extract the overall $(1-z)$-dependence of the master integrals. 

It follows that in the soft limit, each integral scales as $(1-z)^{n - 3 \eps}$ with an integer $n$ that is integral-dependent. 
Hence, all canonical master integrals should be  free of logarithmic singularities as $z \to 1$, or equivalently as $x \to R_1^\pm$, beyond those that  correspond 
to the expansion of $(1-z)^{-3\eps}$ in powers of $\epsilon$. 
This observation allows us  to impose a regularity condition, which fixes 81 integration constants.

The remaining integration constants are obtained by an explicit computation of ten non-canonical integrals in the limit $z \to 1$.
These integrals read
\begin{align}
 B_1 &= \mathcal{I}_{1,1,1,1,1,0,0,0,0,0,0,0}^{\mathrm{T1}}\big|_{s=1,t=1,z\approx1} 
 = (1-z)^{2-3\eps} \,\big( C_1 + \mathcal{O}(1-z) \big)\,,
 \nonumber \\
 B_2 &= \mathcal{I}_{1,1,1,1,1,1,1,0,0,0,0,1}^{\mathrm{T4}} \big|_{s=1,t=1,z\approx1}
 = (1-z)^{-3\eps} \,\big( C_2 + \mathcal{O}(1-z) \big)\,,
 \nonumber \\
 B_3 &= \mathcal{I}_{1,1,1,1,1,1,0,1,1,0,0,1}^{\mathrm{T4}} \big|_{s=1,t=1,z\approx1}
 = (1-z)^{-3\eps} \,\big( C_3 + \mathcal{O}(1-z) \big)\,,
 \nonumber \\
 B_4 &= \mathcal{I}_{1,1,1,1,1,1,1,1,0,0,0,1}^{\mathrm{T10}} \big|_{s=1,t=1,z\approx1}
 = (1-z)^{-1-3\eps} \,\big( C_4 + \mathcal{O}(1-z) \big)\,,
 \nonumber \\
 B_5 &= \mathcal{I}_{1,1,1,1,1,0,0,1,1,0,1,0}^{\mathrm{T19}} \big|_{s=1,t=1,z\approx1}
 = (1-z)^{-3\eps} \,\big( C_5 + \mathcal{O}(1-z) \big)\,,
 \label{eq:B1_through_B10_scaling} \\
 B_6 &= \mathcal{I}_{1,1,1,1,1,0,1,1,1,1,0,1}^{\mathrm{T20}} \big|_{s=1,t=1,z\approx1}
 = (1-z)^{-1-3\eps} \,\big( C_6 + \mathcal{O}(1-z) \big)\,,
 \nonumber \\
 B_7 &= \mathcal{I}_{1,1,1,1,1,1,0,0,1,0,0,1}^{\mathrm{T30}} \big|_{s=1,t=1,z\approx1}
 = (1-z)^{-3\eps} \,\big( C_7 + \mathcal{O}(1-z) \big)\,,
 \nonumber \\
 B_8 &= \mathcal{I}_{1,1,1,1,1,0,1,0,0,0,0,1}^{\mathrm{T32}} \big|_{s=1,t=1,z\approx1}
 = (1-z)^{-3\eps} \,\big( C_8 + \mathcal{O}(1-z) \big)\,,
 \nonumber \\
 B_9 &= \mathcal{I}_{1,1,1,1,1,1,1,0,1,0,0,1}^{\mathrm{T37}} \big|_{s=1,t=1,z\approx1}
 = (1-z)^{-1-3\eps} \,\big( C_9 + \mathcal{O}(1-z) \big)\,,
 \nonumber \\
 B_{10} &= \mathcal{I}_{1,1,1,1,1,1,0,1,1,0,1,1}^{\mathrm{T46}} \big|_{s=1,t=1,z\approx1}
 = (1-z)^{-2-3\eps} \,\big( C_{10} + \mathcal{O}(1-z) \big)\,.
 \nonumber 
\end{align}
To present the results, it is convenient to extract the common $\eps$-dependent factor,
\begin{align}
C_{i} = 
\left(\frac{\Omega_{d-2}}{(2\pi)^{d-1}}\right)^3 
\frac{\Gamma(1-\eps)^6}{\Gamma(1-3\eps)^2}\,
\widetilde{C}_{i}\,,
\label{eq:CtoCtilde}
\end{align}
where $\Omega_n = 2 \pi^{n/2} / \Gamma(n/2)$.
With this normalization, the constants $\widetilde{C}_{i}$ read, up to weight six,
\begin{align}
 \widetilde{C}_1
 &= \left(\tfrac{1}{(1-3 \epsilon )^2 (2-3 \epsilon )^2}\right) \tfrac{1}{16} \,,
 \nonumber \\
 \widetilde{C}_2
 &=  \left(-\tfrac{3}{32 \epsilon ^4}+\tfrac{\pi ^2}{96 \epsilon ^2}+\tfrac{5 \zeta _3}{16 \epsilon }+\tfrac{31 \pi ^4}{2880}+\left(\tfrac{\pi ^2 \zeta _3}{48}+\tfrac{37 \zeta _5}{16}\right) \epsilon +\left(\tfrac{5 \zeta _3^2}{16}+\tfrac{1111 \pi ^6}{181440}\right) \epsilon ^2+\mathcal{O}\left(\epsilon ^3\right)\right) ,
 \nonumber \\
 \widetilde{C}_3
 &=  \left(\tfrac{1}{32 \epsilon ^4}+\tfrac{\pi ^2}{48 \epsilon ^2}+\tfrac{\zeta _3}{\epsilon }+\tfrac{5 \pi ^4}{96}+\left(\tfrac{\pi ^2 \zeta _3}{6}+\tfrac{129 \zeta _5}{8}\right) \epsilon +\left(4 \zeta _3^2+\tfrac{131 \pi ^6}{2160}\right) \epsilon ^2+\mathcal{O}\left(\epsilon ^3\right)\right) ,
 \nonumber \\
 \widetilde{C}_4
 &=  \left(-\tfrac{9}{16 \epsilon ^4}+\tfrac{3 \pi ^2}{16 \epsilon ^2}+\tfrac{27 \zeta _3}{4 \epsilon }+\tfrac{21 \pi ^4}{80}+\tfrac{135 \zeta _5 \epsilon }{2}+\tfrac{31 \pi ^6 \epsilon ^2}{168}+\mathcal{O}\left(\epsilon ^3\right)\right) ,
 \nonumber \\
 \widetilde{C}_5
 &=  \left(-\tfrac{\pi ^2}{48 \epsilon ^2}-\tfrac{\zeta _3}{\epsilon }-\tfrac{77 \pi ^4}{1440}+\left(-\tfrac{1}{8} \pi ^2 \zeta _3-\tfrac{143 \zeta _5}{8}\right) \epsilon +\left(-3 \zeta _3^2-\tfrac{2137 \pi ^6}{30240}\right) \epsilon ^2+\mathcal{O}\left(\epsilon ^3\right)\right) ,
 \label{eq:result_C1_through_C10} \\
 \widetilde{C}_6
 &=  \left(\tfrac{1}{4 \epsilon ^4}-\tfrac{\pi ^2}{32 \epsilon ^2}-\tfrac{3 \zeta _3}{16 \epsilon }+\tfrac{7 \pi ^4}{160}+\left(\tfrac{\pi ^2 \zeta _3}{2}+\tfrac{393 \zeta _5}{16}\right) \epsilon +\left(15 \zeta _3^2+\tfrac{2063 \pi ^6}{15120}\right) \epsilon ^2+\mathcal{O}\left(\epsilon ^3\right)\right) ,
 \nonumber \\
 \widetilde{C}_7
 &=  \left(-\tfrac{\pi ^2}{96 \epsilon ^2}-\tfrac{7 \zeta _3}{16 \epsilon }-\tfrac{\pi ^4}{48}+\left(-\tfrac{1}{12} \pi ^2 \zeta _3-\tfrac{93 \zeta _5}{16}\right) \epsilon +\left(-\tfrac{7 \zeta _3^2}{4}-\tfrac{107 \pi ^6}{5040}\right) \epsilon ^2+\mathcal{O}\left(\epsilon ^3\right)\right) ,
 \nonumber \\
 \widetilde{C}_8
 &= \left(\tfrac{1}{1-3 \epsilon }\right) \left(-\tfrac{\pi ^2}{48 \epsilon }-\tfrac{7 \zeta _3}{8}-\tfrac{\pi ^4 \epsilon }{24}+\left(-\tfrac{1}{6} \pi ^2 \zeta _3-\tfrac{93 \zeta _5}{8}\right) \epsilon ^2+\left(-\tfrac{7 \zeta _3^2}{2}-\tfrac{107 \pi ^6}{2520}\right) \epsilon ^3+\mathcal{O}\left(\epsilon ^4\right)\right) ,
 \nonumber \\
 \widetilde{C}_9
 &=  \left(-\tfrac{3}{32 \epsilon ^4}+\tfrac{\pi ^2}{16 \epsilon ^2}+\tfrac{15 \zeta _3}{4 \epsilon }+\tfrac{37 \pi ^4}{160}+\left(\tfrac{7 \pi ^2 \zeta _3}{8}+81 \zeta _5\right) \epsilon +\left(\tfrac{147 \zeta _3^2}{8}+\tfrac{173 \pi ^6}{504}\right) \epsilon ^2+\mathcal{O}\left(\epsilon ^3\right)\right) ,
 \nonumber \\
 \widetilde{C}_{10}
 &=  \left(-\tfrac{1}{2 \epsilon ^4}+\tfrac{\pi ^2}{16 \epsilon ^2}+\tfrac{3 \zeta _3}{8 \epsilon }-\tfrac{7 \pi ^4}{80}+\left(-\pi ^2 \zeta _3-\tfrac{393 \zeta _5}{8}\right) \epsilon +\left(-30 \zeta _3^2-\tfrac{2063 \pi ^6}{7560}\right) \epsilon ^2+\mathcal{O}\left(\epsilon ^3\right)\right) .
 \nonumber
\end{align}
In the following we describe the various techniques that we  used for computing these  constants.
We discuss the integrals $B_1$, $B_8$, $B_9$ and $B_{10}$ as representative examples. 
All other integrals can be  obtained in similar ways. 
We stress that all results in \cref{eq:result_C1_through_C10} have been checked with an independent numerical calculation, as explained in \cref{sec:checks}.

\subsection{Boundary integral $B_1$}

The boundary integral $B_1$ is equal to the phase-space volume in the limit $z \to 1$.
The phase-space volume is simple enough to be computed directly, keeping the exact dependence on $s,t,z$ and $\eps$.
The relevant integral is given by
\begin{align}
\mathbb{V} = \mathcal{I}_{1,1,1,1,1,0,0,0,0,0,0,0}^{\mathrm{T1}} =
\left(\prod_{i=1}^{3}\int \frac{d^d k_i \,\delta^+(k_i^2)}{(2\pi)^{d-1}}\right)
\delta\big(k_{123}\cdot p - \tfrac{t}{2z}\big)
\delta\big(k_{123}\cdot\pbar - \tfrac{(1-z)s}{2}\big)\,.
\label{eq:phase-space-v1}
\end{align}
It is convenient to introduce the Sudakov decomposition as in  \cref{eq:Sudakov} for all gluon  momenta.
The change of variables from gluon  momenta components to Sudakov parameters leads to
\begin{align}
\mathbb{V} = 
\left(\prod_{i=1}^{3} \frac{s}{2} \int \frac{d \alpha_i \, d\beta_i \, d^{d-2} k_{i \perp} \delta(\alpha_i \beta_i s - k_{i \perp}^2)}{(2\pi)^{d-1}}\right)
\delta\big(\tfrac{\beta_{123}s}{2} - \tfrac{t}{2z}\big)
\delta\big(\tfrac{\alpha_{123}s}{2} - \tfrac{(1-z)s}{2}\big)\,.
\label{eq:phase-space-v2}
\end{align}
We can easily integrate over $k_{i \perp}$ thanks to the on-shell delta function. We obtain 
\begin{align}
\mathbb{V} = 
\left(\prod_{i=1}^{3} \frac{s}{4} \,\frac{\Omega_{d-2}}{(2\pi)^{d-1}} \int d \alpha_i \, d\beta_i \, (\alpha_i \beta_i s)^{-\eps}\right)
\delta\big(\tfrac{\beta_{123}s}{2} - \tfrac{t}{2z}\big)
\delta\big(\tfrac{\alpha_{123}s}{2} - \tfrac{(1-z)s}{2}\big)\,.
\label{eq:phase-space-v3}
\end{align}
We re-scale the Sudakov parameters $\alpha_i = (1-z) \widetilde{\alpha}_i$ and $\beta_i = t/(sz) \widetilde{\beta}_i$, removing the dependencies on $z$ and $t$.
The six remaining integrations factorize into a product of parametric integrals, each of them of the form
\begin{align}
\int\limits_0^1 dx_1 dx_2 dx_3 \, (x_1 x_2 x_3)^{-\eps} \delta(x_{123} - 1)
&= \frac{\Gamma(1-\eps)^3}{\Gamma(3-3\eps)}\,.
\label{eq:param_int_alpha}
\end{align}
As a result, we obtain
\begin{align}
\mathbb{V} &=
\left(\frac{\Omega_{d-2}}{(2\pi)^{d-1}}\right)^3 
\frac{t^{2-3\eps}}{s} 
\frac{1}{16} \frac{\Gamma(1-\eps)^6}{\Gamma(3-3\eps)^2}
\left(\frac{1-z}{z}\right)^{2-3\eps}\,.
\label{eq:phase-space-v4}
\end{align}
The boundary integral $B_1$ is extracted from this expression via
\begin{align}
B_1 
= \mathbb{V} \big|_{s=1,t=1,z \approx 1} 
= 
\left(\frac{\Omega_{d-2}}{(2\pi)^{d-1}}\right)^3 
\frac{1}{16} \frac{\Gamma(1-\eps)^6}{\Gamma(3-3\eps)^2}
(1-z)^{2-3\eps}\,.
\end{align}
Extracting the $z$-dependence and the common $\eps$-dependent pre-factor, as in \cref{eq:CtoCtilde}, we find 
\begin{align}
B_1 = 
(1-z)^{2-3\eps} \,
\left(\frac{\Omega_{d-2}}{(2\pi)^{d-1}}\right)^3 \frac{\Gamma(1-\eps)^6}{\Gamma(1-3\eps)^2} \,
\widetilde{C}_{1} \,,
\end{align}
where 
\begin{align}
\widetilde{C}_{1} = \frac{1}{(2-3\eps)^2\,(1-3\eps)^2} \frac{1}{16}
\end{align}
is the integration constant quoted in \cref{eq:result_C1_through_C10}.

\subsection{Boundary integral $B_8$}

Another relatively simple example is the boundary integral $B_8$, which contains two additional propagators compared to $B_1$. 
Its integral representation reads
\begin{align}
B_8 = 
\left(\prod_{i=1}^{3}\int \frac{d^d k_i \delta^+(k_i^2)}{(2\pi)^{d-1}}\right)
\frac{\delta\big(k_{123}\cdot p - \tfrac{1}{2}\big)\delta\big(k_{123}\cdot\pbar - \tfrac{(1-z)}{2}\big)}
{k_{13}^2 \left(k_{12}\cdot\bar{p}\right)}
\,.
\label{eq:B8v1}
\end{align}
A Sudakov decomposition of the gluon momenta would lead in this case to a non-trivial dependence on the angle between $k_1$ and $k_3$ through the propagator $1/k_{13}^2$.
This situation can be avoided, at least for some of the boundary integrals, by introducing an auxiliary momentum $Q$ that has the effect of factoring out an ordinary phase-space integral.

In the case of $B_8$, it is convenient to choose $Q = k_{13}$ and write
\begin{align}
B_8 &= 
\int \frac{d^d Q}{Q^2}
\int \frac{d^d k_2 \,\delta^+(k_2^2)}{(2\pi)^{d-1}}
\delta\big((Q+k_2)\cdot p - \tfrac{1}{2}\big)\delta\big((Q+k_2)\cdot\pbar - \tfrac{(1-z)}{2}\big)
\widetilde{B}_8(Q^2, Q\cdot\pbar, k_2\cdot\pbar)
\,,
\label{eq:B8v2}
\end{align}
where $\widetilde{B}_8(Q^2, Q\cdot\pbar, k_2 \cdot \pbar)$ is the following integral
\begin{align}
\begin{aligned}
\widetilde{B}_8(Q^2, Q\cdot\pbar, k_2 \cdot \pbar)
&=
\int \frac{d^d k_1 \,\delta^+(k_1^2)}{(2\pi)^{d-1}}
\frac{d^d k_3 \,\delta^+(k_3^2)}{(2\pi)^{d-1}}
\frac{\delta^d\left(Q-k_{13}\right)}{k_{1}\cdot\bar{p} + k_2 \cdot \pbar}
\\
&=
\frac{\Omega_{d-2}}{(2\pi)^{2d-2}} 
\frac{\Gamma^2(1-\eps)}{\Gamma(2-2\eps)}
\frac{\big(Q^2\big)^{-\eps}}{4 k_2 \cdot \pbar}
{}_2F_1\bigg(1,1-\eps;2-2\eps;- \frac{Q\cdot\pbar}{k_2 \cdot \pbar}\bigg)\,.
\label{eq:result_PS_1}
\end{aligned}
\end{align}
The result in \cref{eq:result_PS_1} is most easily obtained by computing the integral in the rest frame of the vector $Q = \big(Q_0, \vec{0}\,\big)$ and expressing the result of the integration  in the Lorentz-invariant way by replacing $Q_0   \pbar_0 $ with $Q \cdot \pbar$ and $Q_0^2$ with $ Q^2$.
Upon inserting the result for the integral into \cref{eq:B8v2}, one can proceed by introducing the Sudakov decomposition for the remaining momenta $k_2$ and $Q$.
Carrying out the resulting parametric integrations yields the desired result
\begin{align}
B_8 &= 
\frac{(1-z)^{-3\eps}}{8}
\left(\frac{\Omega_{d-2}}{(2\pi)^{d-1}}\right)^3
\frac{\Gam^5(1-\eps)\Gam(1-2\eps)\Gam(-\eps)}{\Gam(2-2\eps)\Gam^2(2-3\eps)}
\,{}_3F_2(1,1-\eps,1-2\eps ; 2-2\eps, 2-3\eps ; 1) \,.
\label{eq:B8v3}
\end{align}

\subsection{Boundary integral $B_9$}

It is not always possible to avoid non-trivial angular integrations as in the previous example; this happens in the integrals with multiple propagators of the type $1/k_{ij}^2$. 
As an example, we consider the following boundary integral
\begin{align}
B_9 = 
\left(\prod_{i=1}^{3}\int \frac{d^d k_i \,\delta^+(k_i^2)}{(2\pi)^{d-1}}\right)
\frac{\delta\big(k_{123}\cdot p - \tfrac{1}{2}\big)\delta\big(k_{123}\cdot\pbar - \tfrac{(1-z)}{2}\big)}
{k_{12}^2 \, k_{13}^2 \, \left(k_{13}-p\right)^2 \, \left(k_{12}\cdot\bar{p}\right)} \,.
\label{eq:B9v1}
\end{align}
To calculate it, we use the Sudakov decomposition for each of the gluon momenta $k_i$, c.f. \cref{eq:Sudakov}. 
We then remove the on-shell delta functions $\delta(k_i^2)$ by integrating over $|k_{i,\perp}|$.
Upon re-scaling $\alpha_i \to (1-z) \alpha_i$, we obtain an overall factor $(1-z)^{-1-3\eps}$ while at the same time the parameters $\alpha_i$ become constrained by $\delta(\alpha_{123}-1)$ and are thus placed on an equal footing with the $\beta$-parameters.

Although  the on-shell delta function $\delta(k_i^2)$ fixes the length of the  vector $\vec k_{i\perp}$, its direction remains arbitrary and has to be integrated over. 
The required angular integrations are non-trivial.  
For example, the propagator $1/k_{12}^2$ leads to an angular integral 
\begin{align}
\int \frac{d\Omega_{d-2}^{(2)} }{\alpha_1 \beta_2 + \alpha_2 \beta_1 - 2 \sqrt{\alpha_1 \alpha_2 \beta_1 \beta_2 } \cos\varphi_{12}} 
&= 
\Omega_{d-3} \int\limits_{0}^{\pi} \frac{d\varphi_{12} (1-\cos^2\varphi_{12})^{-\eps}}{\alpha_1 \beta_2 + \alpha_2 \beta_1 - 2 \sqrt{\alpha_1 \beta_2 \alpha_2 \beta_1} \cos\varphi_{12}} 
\nonumber \\[1mm]
&=
\Omega_{d-2} \,
F\left(\alpha_1 \beta_2 , \alpha_2 \beta_1\right)
\,,
\end{align}
where the function $F(x,y)$ reads
\begin{align}
F(x,y) =
\frac{{}_2F_1\left(1+\eps,\tfrac{1}{2}-\eps;1-2\eps;\frac{4 \sqrt{x y}}{\left(\sqrt{x}+\sqrt{y}\right)^2}\right)}{\left(\sqrt{x}+\sqrt{y}\right)^{2}} \,.
\label{eq:defF}
\end{align}
Note that this function is symmetric, i.e. $F(x,y)=F(y,x)$.
The propagator $1/k_{13}^2$ produces a similar function upon integration over the directions of $\vec k_{3\perp}$.
As a result, we obtain
\begin{align}
B_9 =
(1-z)^{-1-3\eps}
\left(\frac{\Omega_{d-2}}{(2\pi)^{d-1}}\right)^{3}  
\left(-\frac{1}{8}\right)
X_9\,,
\label{eq:B9v2}
\end{align}
where the parametric integral $X_9$ is given by
\begin{align}
X_9 &= 
\left(\prod_{i=1}^{3}\int\limits_0^1 d \alpha_i d\beta_i \, (\alpha_i \beta_i)^{-\eps}\right)
\frac{\delta(\alpha_{123}-1)\delta(\beta_{123}-1)}
{\alpha_{12} \, \beta_{13}} 
F\left(\alpha_2 \beta_1, \alpha_1 \beta_2\right)
F\left(\alpha_3 \beta_1, \alpha_1 \beta_3\right)
\,.
\label{eq:X9v1}
\end{align}

We can use the transformation~\cite{abramowitzstegun} 
\begin{align}
{}_2F_1\left(a, b, 2 b, \frac{4 z}{(1 + z)^2}\right) 
= (1 + z)^{2 a} {}_2F_1\left(a, a - b + \tfrac{1}{2}, b + \tfrac{1}{2}, z^2\right)\,, 
\quad |z| < 1 \,,
\label{eq:F21transf}
\end{align}
that simplifies the argument of the hypergeometric function in \cref{eq:defF}. 
We find 
\begin{align}
F(x,y) = \begin{cases}
\frac{{}_2F_1\left(1,1+\eps;1-\eps;\frac{x}{y}\right)}{y} \,, \quad\text{for}\quad x<y\,,\\
\frac{{}_2F_1\left(1,1+\eps;1-\eps;\frac{y}{x}\right)}{x} \,, \quad\text{for}\quad y<x\,.
\end{cases}
\label{eq:FvA}
\end{align}
Since the transformation \cref{eq:F21transf} is only valid if the argument of the hypergeometric function is smaller than one, we must split the integration region into four pieces, according to the cases $\alpha_2 \beta_1 \lessgtr \alpha_1 \beta_2$ and $\alpha_3 \beta_1 \lessgtr \alpha_1 \beta_3$.
Due to the symmetry of the integrand under the simultaneous interchange of subscripts $2 \leftrightarrow 3$ and  $\alpha \leftrightarrow \beta$, two of these contributions happen to be identical.
The calculation of the remaining two contributions is quite similar, so that it is sufficient to describe  the calculation of one of them.

Consider the contribution to $X_9$ that originates from the integration region defined by the conditions $\alpha_2 \beta_1 > \alpha_1 \beta_2$ and $\alpha_3 \beta_1 < \alpha_1 \beta_3$; we will call it $X_9^{(a)}$.
After applying the transformations in \cref{eq:FvA}, 
we find
\begin{align}
X_9^{(a)} &= 
\left(\prod_{i=1}^{3}\int\limits_0^1 d \alpha_i d\beta_i \, (\alpha_i \beta_i)^{-\eps}\right)
\frac{\delta(\alpha_{123}-1)\delta(\beta_{123}-1)}
{\alpha_{12} \, \beta_{13}\,\alpha_1\,\alpha_2 \,\beta_1\,\beta_3} 
\theta(\alpha_2 \beta_1 - \alpha_1 \beta_2)
\label{eq:X9av1} \\&\qquad
\times\theta(\alpha_1 \beta_3 - \alpha_3 \beta_1)\,
{}_2F_1\left(1,1+\eps;1-\eps;\frac{\alpha_1 \beta_2}{\alpha_2 \beta_1}\right)
{}_2F_1\left(1,1+\eps;1-\eps;\frac{\alpha_3 \beta_1}{\alpha_1 \beta_3}\right) \,.
\nonumber
\end{align}
Upon changing variables  $\beta_2 \to r = \alpha_1 \beta_2/(\alpha_2 \beta_1)$ and  $\beta_3 \to \mu = \alpha_3 \beta_1/(\alpha_1 \beta_3)$  and integrating over $\beta_1$ to remove the delta function, we obtain
\begin{align}
\begin{aligned}
X_9^{(a)} &= 
\int \limits_0^1 \frac{dr d\mu \, r^{-\eps}\mu^{-2\eps}}{(1-r)^{1+2\eps}(1-\mu)^{1+2\eps}}
{}_2F_1\left(-2\eps,-\eps;1-\eps;r\right)
{}_2F_1\left(-2\eps,-\eps;1-\eps;\mu\right) 
\\&\quad
\times\left(\prod_{i=1}^{3}\int \limits_0^1 d \alpha_i \, \alpha_i^{-\eps}\right)
\frac{\delta(\alpha_{123}-1)}{\alpha_{12}\,\alpha_1\,\alpha_2} 
(\alpha_3+\alpha_1\mu+\alpha_2 r \mu)^{3\eps}\,
\,.
\label{eq:X9av2}
\end{aligned}
\end{align}
In \cref{eq:X9av2} we have re-written the hypergeometric functions to make them regular in the $r \to 1$ and $\mu \to 1$ limits. 
We proceed by integrating out $\alpha_2$ and change the integration variables $\alpha_1 \to \xi = \alpha_1/(1-\alpha_3)$ and $\alpha_3 \to f = \alpha_3/( \mu (1-\alpha_3))$. 
We obtain 
\begin{align}
\begin{aligned}
X_9^{(a)} &= 
\int \limits_0^1 \frac{dr d\mu \, r^{-\eps}\mu^{-\eps}}{(1-r)^{1+2\eps}(1-\mu)^{1+2\eps}}
{}_2F_1\left(-2\eps,-\eps;1-\eps;r\right)
{}_2F_1\left(-2\eps,-\eps;1-\eps;\mu\right)
\\&\quad
\times\int \limits_0^1 \frac{d\xi \, (1-\xi)^{-2\eps}}{\xi^{1+2\eps}} 
\int \limits_0^\infty \frac{df \, f^{-2\eps} (1+\mu f)^{3\eps}}{f+\xi} 
(f + r + \xi - \xi r)^{3\eps}\,.
\label{eq:X9av3}
\end{aligned}
\end{align}

The integral in \cref{eq:X9av3} is singular; the overlapping logarithmic singularities appear at $r = 1, \mu = 1, \xi = 0$ and $f\in\{0,\infty\}$. 
These singularities are disentangled by performing suitable (iterated) subtractions, after which the resulting integrals are carried out using the program \texttt{HyperInt} \cite{Panzer:2014caa}. 
The other independent contributions are obtained in a similar fashion. 
Upon adding all the contributions, we obtain the result for $X_9$,
\begin{align}
X_9
&= 
\frac{3}{4 \epsilon ^4}
-\frac{5 \pi ^2}{4 \epsilon ^2}
-\frac{42 \zeta _3}{\epsilon }
-\frac{13 \pi ^4}{10}
+\left(43 \pi ^2 \zeta _3-720 \zeta _5\right) \epsilon 
+\left(429 \zeta _3^2-\frac{129 \pi ^6}{140}\right) \epsilon ^2
+ \mathcal{O}(\eps^3)\,.
\label{eq:B9_X9}
\end{align}
The boundary constant $\widetilde{C}_9$ is easily obtained from this result.

\subsection{Boundary integral $B_{10}$} 

The most challenging boundary integrals involve the propagator $1/k_{123}^2$.
Their computation requires a different approach because the Sudakov decomposition of the gluon momenta does not sufficiently simplify them.
To compute these integrals, we set up additional differential equations for suitable parts of their integrands, and determine the boundary constants from these differential equations.
As an example, we consider the last boundary integral
\begin{align}
B_{10} = 
\left(\prod_{i=1}^{3}\int \frac{d^d k_i \,\delta^+(k_i^2)}{(2\pi)^{d-1}}\right)
\frac{\delta\big(k_{123}\cdot p - \tfrac{1}{2}\big)\delta\big(k_{123}\cdot\pbar - \tfrac{(1-z)}{2}\big)}
{k_{123}^2 \, k_{12}^2 \, \left(k_2-p\right)^2 \, \left(k_3\cdot\bar{p}\right) \left(k_{13}\cdot\bar{p}\right)} \,.
\label{eq:B10v1}
\end{align}
The $z$-dependence is again extracted by the re-scaling $k_i \to k_i \sqrt{1-z} $, $\pbar \to \pbar \sqrt{1-z}$ and $p \to p/\sqrt{1-z}$.
We introduce $1 = \int d^dQ \,\delta^d(Q-k_{123})$ and integrate out the momentum $k_3$ to obtain
\begin{align}
B_{10} = 
\frac{(1-z)^{-2-3\eps}}{(2\pi)^{3d-3}}
(-2)
\int d^dQ \, \frac{\delta(2 Q \cdot p - 1) \delta(2 Q \cdot \pbar - 1)}{Q^2}
F_{10}(Q^2)\,.
\label{eq:B10v2}
\end{align}
In \cref{eq:B10v2} we introduced the integral $F_{10}$,
\begin{align}
F_{10}(Q^2) =
\int d^d k_1 d^d k_2 
\frac{\delta^+(k_1^2) \delta^+(k_2^2) \delta^+\big((Q-k_{12})^2\big)}
{k_{12}^2 \, \left(k_2\cdot p\right) \left((Q-k_{12}) \cdot \pbar \right) \left((Q-k_2) \cdot \pbar\right)} \,,
\label{eq:F10v1}
\end{align}
that we will explicitly compute.
As we indicated in \cref{eq:F10v1}, $F_{10}$ depends only on $Q^2$ since all other kinematic invariants are fixed, c.f. \cref{eq:B10v2}.
The variable $Q^2$ satisfies the constraint $0 \leq Q^2 \leq 1$. 
Indeed, the lower boundary appears because $Q^2 = k_{123}^2 \geq 0$, while a Sudakov decomposition of the momentum $Q$ gives $Q^2 = 1 - Q_\perp^2 \leq 1$.

The computation of $F_{10}$ proceeds through the method of differential equations.
We take the derivative of $F_{10}$ with respect to $Q^2$, at fixed $Q \cdot p$ and 
$Q \cdot \bar p$, and, after promoting the delta functions to cut propagators and performing an integration-by-parts reduction, we write the result in terms of a set of masters integrals.
Performing the same steps for the other integrals that contribute to $dF_{10}(Q^2)/dQ^2$, we arrive at a closed system of differential equations that contains five master integrals. 
They are 
\begin{align}
 J_1(Q^2) &= 
 \int d^d k_1 d^d k_2 \,\delta^+(k_1^2) \delta^+(k_2^2) \delta^+\big((Q-k_{12})^2\big)\,,
 \nonumber \\
 J_2(Q^2) &= 
 \int d^d k_1 d^d k_2 \,\frac{\delta^+(k_1^2) \delta^+(k_2^2) \delta^+\big((Q-k_{12})^2\big)}
 {k_{12}^2 \, \left((Q-k_2)\cdot \pbar\right)} \,,
 \nonumber \\
 J_3(Q^2) &= 
  \int d^d k_1 d^d k_2 \,\frac{\delta^+(k_1^2) \delta^+(k_2^2) \delta^+\big((Q-k_{12})^2\big)}
 { \left(k_2 \cdot p\right) \left((Q-k_2)\cdot \pbar\right)} \,,
 \\
 J_4(Q^2) &= 
  \int d^d k_1 d^d k_2 \,\frac{\delta^+(k_1^2) \delta^+(k_2^2) \delta^+\big((Q-k_{12})^2\big)}
 {k_{12}^2 \, \left(k_2 \cdot p\right) \left((Q-k_2)\cdot \pbar\right)}  \,,
 \nonumber \\
  J_5(Q^2) &= F_{10}(Q^2) \,.
  \nonumber 
\end{align}

The differential equations for these master integrals can be easily solved,  but five integration constants need to be determined.
We obtain these integration constants by various means.
One constant follows from the calculation of $J_1(Q^2)$ at $Q^2 = 1$, which is closely related to the phase-space integral $B_1$.
Constraints on the remaining integration constants are obtained from the analysis of the solutions to the differential equations in the limits $Q^2 \to 0$ and $Q^2 \to 1$. 
For example, we require that $J_5(Q^2) = F_{10}(Q^2)$ does not have a hard region $(Q^2)^0$, because the integral in \cref{eq:B10v2} would otherwise be ill-defined.
We also find that the $(Q^2)^0$ branch of $J_3(Q^2)$ vanishes, that the $(Q^2)^{-\eps}$ branch of $J_4(Q^2)$ vanishes and that the $(1-Q^2)^0$ branch of $J_2(Q^2)$ is given by
\begin{align}
J_2\big|_{Q^2=1} = 
\frac{\Omega_{d-2}^2}{8} \,
\frac{\Gamma^4(1 - \eps) \Gamma( - \eps)}{\Gamma(2 - 3 \eps) \Gamma(2 - 2 \eps)} \,
{}_3F_2\left(1, 1 - 2 \eps, 1 - \eps ; 2 - 3 \eps , 2 - 2 \eps ; 1\right)\,.
\end{align}
Putting all this information together gives us the result for $F_{10}(Q^2)$.
Using it in \cref{eq:B10v2} and integrating over $Q$, we obtain the boundary integral $B_{10}$.
It reads
\begin{align}
B_{10} = 
(1-z)^{-2-3\eps}
\left(\frac{\Omega_{d-2}^3}{(2\pi)^{d-1}}\right)^3
X_{10} \,,
\label{eq:B10vXX}
\end{align}
where $X_{10}$ is given by the following expression
\begin{align}
X_{10}
&= 
- \frac{1}{2 \eps^4} 
+ \frac{9 \pi^2}{16 \eps^2} 
+ \frac{67 \zeta_3}{8 \eps}
- \frac{11 \pi^4}{60}
- \left(\frac{83}{8} \pi^2 \zeta_3 + \frac{9}{8} \zeta_5\right) \eps 
- \left(\frac{575}{6}  \zeta_3^2 + \frac{445}{2268} \pi^6\right) \eps^2
+ \mathcal{O}(\eps^3)\,.
\label{eq:B10X10}
\end{align}
The constant $\widetilde{C}_{10}$ is then easily extracted.


\section{Numerical checks of master integrals}
\label{sec:checks}

We have performed several checks to ensure the correctness of the master integrals computed in the previous section.
First, we inserted the master integrals into the system of differential equations from which they were derived and checked that the differential equations are indeed satisfied. 
Second, some of the boundary constants for $z\rightarrow 1$ have been computed in several different ways. 
Nevertheless, a completely independent check of the integrals is desirable. 
Unfortunately, contrary to standard Feynman integrals, there exists no automated code to evaluate phase-space integrals numerically and therefore we have to proceed differently.

In Ref.~\cite{Melnikov:2018jxb} we have considered similar phase-space integrals, albeit in a situation where one of the gluons was virtual and two were real.
The double-virtual single-real master integrals  in that paper were checked numerically using the Mellin-Barnes (MB) integration, following the discussion in  Ref.~\cite{Anastasiou:2013srw}. 
We employ the same approach to check the triple-real integrals computed in the current paper; since there are significant similarities with the calculation described in Ref.~\cite{Melnikov:2018jxb}, we only give a short overview of the steps required for the numerical checks.

For reasons explained in Ref.~\cite{Melnikov:2018jxb}, in order to perform the numerical evaluation of the phase-space integrals, it is preferable to consider the {\it decay} process $q^* \to q+ggg$ instead of the production process $q \to q^* + ggg$. 
We accomplish this by formally changing the four-momenta $p\rightarrow -p, \bar{p}\rightarrow -\bar{p}$ in the definition of the master integrals. 
We obtain
\begin{align}
\begin{aligned}
\delta(k_{123}\cdot \bar{p} - (1-z)/2) \,\delta(k_{123}\cdot p - \kappa/2)& \longrightarrow \delta(k_{123}\cdot \bar{p} + (1-z)/2) \,\delta(k_{123}\cdot p + \kappa/2) \,,\\
 (p-k_{i\ldots j})^2 & \longrightarrow (p+k_{i\ldots j})^2\,, 
\end{aligned}
\end{align}
where we introduced $\kappa=t/z$. 
It follows from the above equation that we need to take $z \geq  1$ and $t\leq 0$, or otherwise the integrals would identically
 vanish. 
The analytic expression for the integral in the decay kinematics, that we refer to as $I_{\text{decay}}(\kappa,z)$, can be determined from the solutions in the production channel $I_{\text{production}}(\kappa,z)$ by an analytic continuation to the region $z \geq  1$ and $\kappa\leq 0$. 
Note that since the propagators are positive definite both in the production and in the decay kinematics, both integrals $I_{\text{production}}(\kappa,z)$ and $I_{\text{decay}}(\kappa,z)$ are {\it real-valued}. 
This consideration provides a useful constraint on the results of the analytic continuation. 

As the next step, we set $\kappa=z-2$ and write 
\begin{equation}
W = \int\limits_1^2 dz \, I_{\text{decay}}(z-2,z) = \int\limits_{-\infty}^{\infty} d\kappa \int\limits_{-\infty}^{\infty} dz \, I_{\text{decay}}(\kappa,z)\, \delta(z-2-\kappa).
\label{eq:check}
\end{equation}
Note that in the second step in \cref{eq:check} we used the fact that $I_{\text{decay}}(\kappa,z)$ vanishes outside the region $\kappa\leq 0, z\geq 1$.

\Cref{eq:check} can be used to check our integrals numerically. 
Indeed, on the one hand, the first integral in \cref{eq:check} can be calculated  
using the 
analytic solution $I_{\text{production}}(\kappa,z)$, continued to the decay region $z \geq  1$, $\kappa\leq 0$. 
On the other hand, $W$ can be written as a MB integral, following the discussion in Ref.~\cite{Anastasiou:2013srw}.
Indeed, we consider the right-hand side of \cref{eq:check} and write the integral as 
\begin{align}
\begin{split}
W & = 
\int \limits_{-\infty}^{\infty} d\kappa \int \limits_{-\infty}^{\infty} dz \int 
\prod_{i=1}^3 \left[ {d}^d k_i\delta^{+}(k_i^2) \right] \delta(z-2-\kappa) \, \delta\Big(k_{123}\cdot \bar{p} + \frac{1-z}{2}\Big)\, \delta\Big(k_{123}\cdot p + \frac{\kappa}{2}\Big) \prod_j \frac{1}{D_j} \\
& = 4
\int
\prod_{i=1}^3 {d}^d k_i \delta^{+}(k_i^2) \,\delta(1-2k_{123}\cdot (p+\bar{p})) \prod_j \frac{1}{D_j},
\end{split} 
\label{eq:checkW}
\end{align}
where $D_j$ are the propagators of the particular integral, c.f. \cref{tab:inverseprops}.
To proceed further, we may use the Mellin-Barnes representation  
 \begin{gather}
\frac{1}{(x+y)^\lambda}=\int \limits_{-i\infty}^{+i\infty} \frac{dz}{2\pi i} \frac{y^z}{x^{z+\lambda}}\frac{\Gamma(-z)\Gamma(\lambda+z)}{\Gamma(\lambda)}\,,
\label{eq:MB}
\end{gather}
 to re-write propagators of the form 
\begin{equation}
\frac{1}{(p+k_{ij})^2} =
\frac{1}{2 p \cdot k_i + 2 p \cdot k_j + 2 k_i \cdot k_j}
\end{equation}
into integrals of products of $k_i \cdot k_j$, $p \cdot k_i$ and $p \cdot k_j$. 
Upon doing so, we obtain integrals that are identical to the ones studied in Ref.~\cite{Anastasiou:2013srw} and we can follow that reference to construct the Mellin-Barnes representation for those integrals. 
The resulting Mellin-Barnes integrals are finally computed numerically with the package MBtools~\cite{MBTools}. 
The two results for the quantity $W$ in \cref{eq:check} must agree and we, therefore, get an indirect check of the results for the master integrals.
We have performed this comparison for the master integrals and found agreement within the numerical errors. 
Furthermore, we note that we can use the same procedure to compute the soft limits of all integrals, checking the boundary values for all of them through weight six.


\section{Results}
\label{sec:results}

The analytic expressions for the 91 master integrals $\mathcal{I}_{\vec{n}}^{\mathrm{top}}$ listed in \cref{eq:listofmasters} are the main result of this paper.
To present them we choose the normalization such that
\begin{align}
\mathcal{I}_{\vec{n}}^{\mathrm{top}} 
=
s^{-\mathcal{M}} \, t^{\mathcal{N}}
\left(\frac{\Omega_{d-2}}{(2\pi)^{d-1}}\right)^3 
\frac{\Gamma(1-\eps)^6}{\Gamma(3-3\eps)^2}
\left(\frac{1-z}{z}\right)^{2-3\eps}
{\rm INT}(\mathrm{top},\vec{n}) 
\,,
\end{align}
where the powers $\mathcal{M}$ and $\mathcal{N}$ depend on the index vector $\vec{n}$, as explained in \cref{eq:after_scaling_st}.
With this normalization, the integral related to the phase-space volume becomes
\begin{align}
{\rm INT}(\mathrm{T1}, \{1,1,1,1,1,0,0,0,0,0,0,0\}) = \frac{1}{16}\,.
\end{align}
In general, the integrals ${\rm INT}(\mathrm{top},\vec{n})$ depend on the variable $x$, which is related to the longitudinal momentum fraction $z$ via \cref{eq:LandauVariable}.
We did not express all the master integrals in terms of the variable $z$ since, if one does this, square roots of $z$ appear. 
Explicit expressions for the integrals ${\rm INT}(\mathrm{top},\vec{n})$ are provided in an ancillary file, which may be downloaded from \url{https://www.ttp.kit.edu/_media/progdata/2019/ttp19-009.tar.gz}.

To illustrate the usefulness of the integrals presented in this paper, we construct the RRR contribution to the $I_{qq}$ matching coefficient at N${}^3$LO in QCD in the large-$N_c$ and the large-$N_f$ limits.
Interestingly, upon inserting our results for the master integrals, we find that all $x$-dependent multiple polylogarithms as well as the rational functions of $x$ combine in such a way, that the final result is expressible in terms of rational functions of $z$ and harmonic polylogarithms of $z$ only. 
The required mappings from $G(\vec{w}; x)$ to $H(\vec{w},z)$ were obtained by expressing all harmonic polylogarithms up to weight $6$ in terms of $G(\vec{w}; x)$ with the program \texttt{HyperInt} \cite{Panzer:2014caa} and subsequently inverting the (underdetermined) system of linear equations.
The resulting $s$-independent contributions can be written as
\begin{align}
\mathcal{A}_i(t,z,\eps)
=
g_s^6
\left(\frac{\Omega_{d-2}}{(2\pi)^{d-1}}\right)^3 
\frac{\Gamma(1-\eps)^6}{\Gamma(3-3\eps)^2}\,
t^{-1-3\eps}
\mathcal{A}_i(z,\eps) \,,
\end{align}
where $g_s$ is the strong coupling constant. 
The subscript $i$ is either $N_c^3$ to indicate the leading-color contribution, or $N_f$ to indicate the contribution proportional to $N_f$.
The $t$-dependence factorizes by construction, since we computed the leading contribution in the collinear limit.
In fact, this factor will eventually be expanded in terms of plus distributions,
\begin{align}
t^{-1+k\eps}
= \frac{\delta(t)}{k \eps} + \sum_{n \geq 0} \frac{(k\eps)^n}{n!} \left[\frac{\log^n (t)}{t}\right]_{+}\,,
\label{eq:tdistrib}
\end{align}
in order to properly extract the collinear singularities.
As a consequence, $\mathcal{A}_i(z,\eps)$ is needed up to first order in $\eps$.
In turn, $\mathcal{A}_i(z,\eps)$ contains soft singularities, which are extracted by writing $(1-z)^{-1-3\eps}$ in terms of plus distributions.
The results are rather lengthy, so we choose to only display their soft limits.
They read
\begin{align}
\mathcal{A}_{N_c^3}(z,\eps) &= N_c^3 \delta (1-z) 
\Bigg(\!\!
-\frac{100}{3 \epsilon ^5}
+\frac{724}{3 \epsilon ^4}
+\frac{1}{\epsilon ^3} \bigg(\!\! -\frac{5471}{9}+\frac{17 \pi ^2}{3}\bigg)
\nonumber\\&\quad
+\frac{1}{\epsilon ^2} \bigg(230 \zeta _3-\frac{437 \pi ^2}{9}+\frac{5942}{9}\bigg)
+\frac{1}{\epsilon } \bigg(\!\! -\frac{5902 \zeta _3}{3}+\frac{472 \pi ^4}{45}+\frac{16061 \pi ^2}{108}-\frac{20129}{162}\bigg)
\nonumber\\&\quad
+\bigg(
\frac{1651}{486}
-\frac{15806 \pi ^2}{81}
+\frac{108215 \zeta _3}{18}
-\frac{4028 \pi ^4}{45}
+20 \pi ^2 \zeta _3
+3042 \zeta _5
\bigg)
\nonumber\\&\quad
+\bigg(\!\!
-\frac{9116}{81}
+\frac{3448181 \pi ^2}{62208}
-\frac{212752 \zeta _3}{27}
+\frac{36818 \pi ^4}{135}
-\frac{1444 \pi ^2 \zeta _3}{9}
-26014 \zeta _5
\nonumber\\&\quad
+384 \zeta _3^2
+\frac{1999 \pi ^6}{189}
\bigg) \epsilon 
+O\big(\epsilon ^2\big)
\Bigg)\,,
\\
\mathcal{A}_{N_f}(z,\eps)
&=
C_F^2 N_f \delta (1-z) \Bigg(
\frac{44}{9 \epsilon ^4}
-\frac{932}{27 \epsilon ^3}
+\frac{6425}{81 \epsilon ^2}
-\frac{15203}{243 \epsilon }
+\frac{5315}{729}
+\frac{8443}{2187}\epsilon
+\mathcal{O}\big(\epsilon ^2\big)
\Bigg)
\nonumber\\&\quad
+C_A C_F N_f \delta (1-z) \Bigg(
\frac{2}{3 \epsilon ^4}
-\frac{62}{27 \epsilon ^3}
-\frac{1}{\epsilon ^2}\bigg(\frac{133}{162}+\frac{4 \pi ^2}{27}\bigg)
\nonumber\\&\quad
+\frac{1}{\epsilon }\bigg(\frac{158}{27}+\frac{88 \pi ^2}{81}
-\frac{56 \zeta_3}{9}\bigg)
+\bigg(\!\!
-\frac{7060}{729}-\frac{427 \pi ^2}{243}+\frac{1232 \zeta_3}{27}-\frac{8 \pi ^4}{27}
\bigg)
\nonumber\\&\quad
+\bigg(
\frac{43033}{4374}
-\frac{2501 \pi ^2}{729}
-\frac{5762 \zeta _3}{81}
+\frac{176 \pi ^4}{81}
-\frac{32 \pi ^2 \zeta_3}{27}
-\frac{248 \zeta_5}{3}\bigg) \epsilon 
+O\big(\epsilon ^2\big)
\Bigg)\,.
\nonumber
\end{align}


\section{Conclusions} 
\label{sec:concl}

In this paper, we computed  the master integrals  required to describe  the  real-emission contribution to the matching coefficient of a quark beam function at N$^3$LO in QCD
due to the splitting of an incoming quark $q$ into a virtual quark of the same flavor and three collinear gluons, $q \to q^*+ggg$. 
We  used reverse unitarity and integration-by-parts identities to derive  differential equations satisfied by the master integrals.  
We solved the differential equations and fixed  the  boundary conditions for the master integrals using  both 
regularity requirements  and the explicit computation of a small subset of integrals in the soft limit.
Our final results for the master integrals are expressed in terms of Goncharov  polylogarithms up to weight six.

The master integrals computed in this paper allow us to obtain the 
triple-real contribution to the matching coefficient $I_{qq}$ in the large-$N_c$ and large-$N_f$ limits.   
To extend this calculation to include terms that are sub-leading  in $N_c$, we have to account for processes where an incoming quark $q$ splits into a quark-antiquark pair of the same  flavor and a gluon, $q \to q^* + q \bar{q} g$.
The contribution  of  this process to $I_{qq}$  requires additional master integrals. We expect their computation to be feasible using the techniques described in this paper. 

As we pointed out in the Introduction, there are three N$^3$LO  QCD contributions to $I_{qq}$, the triple-real, the double-real single-virtual and the single-real double-virtual, that need to be calculated. 
We studied the double-real single-virtual contribution in Ref.~\cite{Melnikov:2018jxb} and the  triple-real contribution in this  paper. 
The so far unattended contribution is the single-real double-virtual one; its computation will require us to understand how to compute a massless   two-loop three-point function in an axial gauge. 
Although such a computation appears to be quite challenging, we believe that it can be dealt with  using calculational methods developed both in this paper and in Ref.~\cite{Melnikov:2018jxb}.

\appendix
\crefalias{section}{appsec}

\section{List of master integrals}
\label{app:listofmasters}

In this appendix we list the $91$ master integrals.
\begin{align}
&\mathcal{I}^{A_1}_{0, 0, 0, 0, 0, 0, 0} \,,\;\;
\mathcal{I}^{A_1}_{0, 0, 1, 0, 0, 0, 0} \,,\;\;
\mathcal{I}^{A_1}_{0, 0, 0, 0, 1, 0,0} \,,\;\;
\mathcal{I}^{A_1}_{0, 1, 1, 1, 0, 0, 0} \,,\;\;
\mathcal{I}^{A_1}_{-1, 1, 1, 1, 0, 0, 0} \,,\;\;
\mathcal{I}^{A_1}_{0, 1, 1, 1, -1,0, 0} \,,\;\;  \nonumber \\ &
\mathcal{I}^{A_1}_{0, 1, 1, 1, 0, -1, 0} \,,\;\; 
\mathcal{I}^{A_1}_{0, 0, 1, 0, 0, 1, 0} \,,\;\;
\mathcal{I}^{A_1}_{0, 1, 1,0, 0, 1, 0} \,,\;\;
\mathcal{I}^{A_1}_{1, 0, 0, 0, 1, 1, 0} \,,\;\;
\mathcal{I}^{A_1}_{0, 1, 0, 0, 1, 1, 0} \,,\;\;
\mathcal{I}^{A_1}_{0, 0, 1, 0, 1, 1, 0} \,,\;\; \nonumber  \\ &
\mathcal{I}^{A_1}_{0, 0, 0, 1, 1, 0,1} \,,\;\;
\mathcal{I}^{A_1}_{0, 0, 1, 1, 0, 0, 1} \,,\;\;  
\mathcal{I}^{A_1}_{0, 0, 1, 2, 0, 0, 1} \,,\;\;
\mathcal{I}^{A_1}_{0, 0, 2, 1, 0,0, 1} \,,\;\;
\mathcal{I}^{A_1}_{0, 1, 0, 1, 1, 1, 0} \,,\;\;
\mathcal{I}^{A_1}_{0, 1, 1, 1, 0, 0, 1} \,,\;\;  \nonumber  \\ &
\mathcal{I}^{A_1}_{1, 0, 1, 0,0, 1, 1} \,,\;\;
\mathcal{I}^{A_1}_{1, 0, 0, 1, 1, 1, 1} \,,\;\;
\mathcal{I}^{A_1}_{1, 0, 1, 0, 1, 1, 1} \,,\;\; 
\mathcal{I}^{A_1}_{1, 1, 0,1, 1, 1, 0} \,,\;\;
\mathcal{I}^{A_2}_{0, 1, 1, 1, 0, 0, 1} \,,\;\;
\mathcal{I}^{A_2}_{-1, 1, 1, 1, 0, 0, 1} \,,\;\; \nonumber  \\ &
\mathcal{I}^{A_2}_{1, 1, 0, 1, 1, 0, 1} \,,\;\;
\mathcal{I}^{A_2}_{0, 1, 1, 1, 1, 0,1} \,,\;\;
\mathcal{I}^{A_2}_{1, 1, 0, 0, 1, 1, 1} \,,\;\;
\mathcal{I}^{A_2}_{0, 1, 1, 0, 1, 1, 1} \,,\;\;  
\mathcal{I}^{A_2}_{1, 1, 0, 1, 1, 1, 1} \,,\;\;
\mathcal{I}^{A_3}_{0, 0, 1, 1, 0, 1, 1} \,,\;\;  \nonumber  \\ &
\mathcal{I}^{A_3}_{0, 1, 1, 1, 0, 1, 1} \,,\;\;
\mathcal{I}^{A_4}_{1, 0, 0, 1,0, 0, 0} \,,\;\;
\mathcal{I}^{A_4}_{1, -1, 0, 1, 0, 0, 0} \,,\;\;
\mathcal{I}^{A_4}_{1, 1, 0, 0, 0, 0, 1} \,,\;\;
\mathcal{I}^{A_4}_{1, 0, 0, 1, 0, 0, 1} \,,\;\; 
\mathcal{I}^{A_4}_{1, 1, 0, 1, 1, 0,0} \,,\;\;  \nonumber \\ &
\mathcal{I}^{A_4}_{1, 1, 0, 1, 0, 0, 1} \,,\;\;
\mathcal{I}^{A_4}_{1, 0, 1, 1, 0, 0, 1} \,,\;\;
\mathcal{I}^{A_4}_{1, -1, 1, 1, 0,0, 1} \,,\;\;
\mathcal{I}^{A_4}_{1, 0, 1, 1, -1, 0, 1} \,,\;\;
\mathcal{I}^{A_4}_{1, 1, 0, 0, 1, 0, 1} \,,\;\;
\mathcal{I}^{A_4}_{1, 1, -1,0, 1, 0, 1} \,,\;\;  \nonumber  \\ &
\mathcal{I}^{A_5}_{1, 0, 0, 1, 0, 0, 1} \,,\;\;
\mathcal{I}^{A_5}_{1, 1, 0, 1, 0, 0, 1} \,,\;\;
\mathcal{I}^{A_5}_{1, 0, 1, 1, 0, 0, 1} \,,\;\;
\mathcal{I}^{A_5}_{1, 1, 0, 1, 1, 0,1} \,,\;\;
\mathcal{I}^{A_5}_{1, 1, -1, 1, 1, 0, 1} \,,\;\;
\mathcal{I}^{A_6}_{1, 0, 0, 1, 0, 1, 1} \,,\;\;  \label{eq:listofmasters} \\ &
\mathcal{I}^{A_6}_{1, 0, 1, 1,0, 1, 1} \,,\;\;  
\mathcal{I}^{A_6}_{1, 1, 0, 0, 1, 1, 1} \,,\;\;
\mathcal{I}^{A_7}_{1, 1, 1, 1, 0, 0, 0} \,,\;\;
\mathcal{I}^{A_7}_{1, 1, 1, 1, 1, 0, 0} \,,\;\;
\mathcal{I}^{A_7}_{1, 1, 1, 1, 0, 1, 1} \,,\;\;
\mathcal{I}^{A_8}_{0, 1, 1, 1, 0, 1, 0} \,,\;\; \nonumber  \\ &
\mathcal{I}^{A_8}_{0, 1, 1, 1, 0, 1, 1} \,,\;\; 
\mathcal{I}^{A_9}_{1, 1, 1, 0, 0, 0,1} \,,\;\;  
\mathcal{I}^{A_9}_{1, 1, 1, 0, 1, 0, 1} \,,\;\;
\mathcal{I}^{A_9}_{1, 1, 1, -1, 1, 0, 1} \,,\;\;
\mathcal{I}^{A_{10}}_{1, 1, 1, 0,1, 0, 1} \,,\;\;
\mathcal{I}^{A_{11}}_{1, 1, 1, 0, 1, 1, 1} \,,\;\;  \nonumber  \\ &
\mathcal{I}^{A_{12}}_{0, 0, 1, 0, 1, 0, 0} \,,\;\;
\mathcal{I}^{A_{12}}_{0, 0, 1, 1, 0, 1, 0} \,,\;\;
\mathcal{I}^{A_{13}}_{0, 0, 1, 0, 1, 0,1} \,,\;\;  
\mathcal{I}^{A_{13}}_{0, 1, 1, 1, 1, 0, 1} \,,\;\;
\mathcal{I}^{A_{13}}_{-1, 1, 1, 1, 1, 0, 1} \,,\;\;
\mathcal{I}^{A_{14}}_{0, 0, 1, 0,0, 0, 1} \,,\;\;   \nonumber \\ &
\mathcal{I}^{A_{14}}_{-1, 0, 1, 0, 0, 0, 1} \,,\;\; 
\mathcal{I}^{A_{14}}_{1, 0, 1, 0, 0, 0, 1} \,,\;\;
\mathcal{I}^{A_{14}}_{1, 0, 0, 0, 1, 0, 1} \,,\;\;
\mathcal{I}^{A_{14}}_{1, 0, -1, 0, 1, 0,1} \,,\;\; 
\mathcal{I}^{A_{14}}_{0, 0, 1, 0, 1, 0, 1} \,,\;\;
\mathcal{I}^{A_{14}}_{-1, 0, 1, 0, 1, 0, 1} \,,\;\;  \nonumber \\ &
\mathcal{I}^{A_{14}}_{1, 0, 1, 0,1, 0, 1} \,,\;\; 
\mathcal{I}^{A_{14}}_{1, 0, 0, 1, 0, 1, 1} \,,\;\;
\mathcal{I}^{A_{14}}_{1, 1, 0, 0, 1, 1, 1} \,,\;\;
\mathcal{I}^{A_{14}}_{1, 0, 1, 0, 1, 1, 1} \,,\;\;
\mathcal{I}^{A_{14}}_{0, 1, 1, 0, 1, 1, 1} \,,\;\;  
\mathcal{I}^{A_{14}}_{1, 0, 0, 1, 1, 1, 1} \,,\;\; \nonumber \\ &
\mathcal{I}^{A_{14}}_{1, 1, 0, 1, 1, 1, 1} \,,\;\; 
\mathcal{I}^{A_{15}}_{1, 0, 0, 1,0, 0, 1} \,,\;\;
\mathcal{I}^{A_{15}}_{1, 1, 0, 1, 0, 0, 1} \,,\;\;
\mathcal{I}^{A_{15}}_{1, 1, 0, 1, 1, 0, 1} \,,\;\;
\mathcal{I}^{A_{15}}_{1, 1, -1, 1, 1, 0, 1} \,,\;\;
\mathcal{I}^{A_{16}}_{0, 1, 0, 0, 0, 0,1} \,,\;\;   \nonumber \\ &
\mathcal{I}^{A_{16}}_{0, 1, 0, 0, 1, 0, 1} \,,\;\;
\mathcal{I}^{A_{17}}_{1, 0, 1, 0, 1, 0, 1} \,,\;\;
\mathcal{I}^{A_{17}}_{1, 0, 1, 0,1, 1, 1} \,,\;\;
\mathcal{I}^{A_{18}}_{1, 1, 0, 1, 0, 0, 1} \,,\;\;
\mathcal{I}^{A_{18}}_{1, 1, 0, 1, 1, 0, 1} \,,\;\;
\mathcal{I}^{A_{18}}_{1, 1, -1, 1, 1, 0, 1} \,,\;\;  \nonumber \\ &
\mathcal{I}^{A_{19}}_{1, 0, 1, 1, 0, 1, 1}\,. \nonumber 
\end{align}
The definition of the topologies $A_1$ through $A_{19}$ may be found in \cref{tab:inverseprops}.

\section{Additional relations among the master integrals}
\label{app:extraids}

In this appendix we prove  two simple relations  among some of the master integrals, \cref{eq:extraidentity1,eq:extraidentity2}.
We define the two quantities
\begin{align}
&F_1(z, \eps) = \mathcal{I}^{A_{10}}_{1,1,1,0,1,0,1} + \frac{1}{2} \frac{(1-z)}{z}\;  \mathcal{I}^{A_{6}}_{1,1,0,0,1,1,1} \,, \\
&F_2(z, \eps) = \mathcal{I}^{A_{1}}_{0,0,1,0,0,1,0} + \frac{2(1-2\eps)}{\eps} \frac{1}{z}\;  \mathcal{I}^{A_{12}}_{0,0,1,0,1,0,0}\,.
\end{align}
Below we show  that $F_{1}(z,\eps) = F_{2}(z,\eps) = 0$.

Using the result for the differential equations for the master integrals, we find that  $F_1$ and $F_2$ satisfy the following homogeneous differential equations
\begin{align}
& \frac{dF_1(z, \eps)}{dz} = \frac{1}{z}\; F_1(z, \eps)
\,,
\\
& \frac{dF_2(z, \eps)}{dz} = \eps \left( \frac{1}{1-z} - \frac{1-3\eps}{z} \right) F_2(z,\eps)
 \,.
\end{align}
The solutions to these equations are 
\begin{align}
& F_1(z, \eps) = c_1(\eps)\, z\,,\;\;\;
F_2(z, \eps) = c_2(\eps)\, z^{-1+3\eps} (1-z)^{-\eps} \,.
\label{eqB5}
\end{align}
In the limit $z \to 1$ these solutions for $F_1$ and $F_2$ behave as $(1-z)^0$ and $(1-z)^{-\eps}$.
However, we have argued in the main body of the paper that all master integrals in the soft $ z \to 1$ limit should be proportional to $(1-z)^{n-3\eps}$ for some integer $n$. The only way to make this scaling compatible with \cref{eqB5} is to choose  
$c_{1}(\eps) = c_{2}(\eps) = 0$ which implies that 
$F_{1,2}$ vanish identically. 
This proves the identities among master integrals shown in \cref{eq:extraidentity1,eq:extraidentity2}.

\acknowledgments
We thank A. von Manteuffel for a useful advice concerning  the calculation of boundary integrals. 
L.T. would like to acknowledge the Mainz Institute for Theoretical Physics (MITP), which is part of the DFG Cluster of Excellence PRISMA$^+$ (Project ID 39083149), for its partial support during the completion of this work.
The research of K.M. and R.R. was supported by the Deutsche Forschungsgemeinschaft (DFG, German Research Foundation) under grant  396021762 - TRR 257. The research of L.T. was supported by the ERC
starting grant 637019 “MathAm”. 
The research of Ch.W. was supported in part by the BMBF project No. 05H18WOCA1.

\bibliographystyle{JHEP}
\bibliography{biblio}
\end{document}